\documentclass[aps,pre,reprint,superscriptaddress]{revtex4-1}

\pdfoutput=1

\usepackage{graphicx}
\usepackage{amssymb}
\usepackage{amsmath}
\usepackage{xcolor}
\usepackage{pdfpages}
\usepackage{pgffor}
\usepackage{CJK}

\makeatletter
\AtBeginDocument{\let\LS@rot\@undefined}
\makeatother

\begin{document}

\begin{CJK*}{UTF8}{} 

\title{Control of Stratification in Drying Particle Suspensions via Temperature Gradients}

\author{Yanfei Tang ({\CJKfamily{gbsn}唐雁飞})}
\affiliation{Department of Physics, Center for Soft Matter and Biological Physics,
and Macromolecules Innovation Institute, Virginia Polytechnic Institute and State University,
Blacksburg, Virginia 24061, USA}
\author{Gary S. Grest}
\affiliation{Sandia National Laboratories, Albuquerque, NM 87185, USA}
\author{Shengfeng Cheng ({\CJKfamily{gbsn}程胜峰})}
\affiliation{Department of Physics, Center for Soft Matter and Biological Physics,
and Macromolecules Innovation Institute, Virginia Polytechnic Institute and State University,
Blacksburg, Virginia 24061, USA}
\email{chengsf@vt.edu}

\date{\today}

\begin{abstract}
A potential strategy for controlling stratification in a drying suspension of bidisperse particles is studied using molecular dynamics simulations. When the suspension is maintained at a constant temperature during fast drying, it can exhibit  ``small-on-top'' stratification with an accumulation (depletion) of smaller (larger) particles in the top region of the drying film, consistent with the prediction of current theories based on diffusiophoresis. However, when only the region near the substrate is thermalized at a constant temperature, a negative temperature gradient develops in the suspension because of evaporative cooling at the liquid-vapor interface. Since the associated thermophoresis is stronger for larger nanoparticles, a higher fraction of larger nanoparticles migrate to the top of the drying film at fast evaporation rates. As a result, stratification is converted to ``large-on-top''. Very strong ``small-on-top'' stratification can be produced with a positive thermal gradient in the drying suspension. Here we explore a way to produce a positive thermal gradient by thermalizing the vapor at a temperature higher than that of the solvent. Possible experimental approaches to realize various thermal gradients in a suspension undergoing solvent evaporation, and thus to produce different stratification states in the drying film, are suggested.
\end{abstract}

\maketitle

\end{CJK*}

\section{INTRODUCTION}

The drying of colloidal suspensions has been studied for several decades.\cite{Sheetz1965, Keddie1997, Steward2000, Russel2011, Routh2013, Zhou2017b, Schulz2018, Cairncross1996, Routh2004, Cardinal2010, Nikiforow2010, Trueman2012, Trueman2012a, Atmuri2012, Cheng2013, Doumenc2016, Howard2018} Recently, drying-induced stratification phenomena in polydisperse colloidal mixtures have attracted great attention,\cite{Fortini2016, Martin-Fabiani2016, Zhou2017, Fortini2017, Howard2017, Howard2017b, Makepeace2017, Sear2017, Sear2018, Liu2018, Tatsumi2018, Cusola2018, Tang2018, Carr2018, Statt2018, Martin-Fabiani2018, Zhou2017b, Schulz2018} as they point to a quick, facile, one-pot method of depositing layered multifunctional coating films on a surface. In a particle suspension undergoing drying, the vertical distribution of particles is controlled by the P\'{e}clet number, $\text{Pe} = Hv_e/D$, where $H$ is the thickness of the suspension film, $v_e$ is the receding speed of the liquid-vapor interface during evaporation, and $D$ is the diffusion coefficient of the particles.\cite{Routh1998, Routh2004} The P\'{e}clet number characterizes the competition between diffusion and evaporation-induced particle migration. When $\text{Pe} \gg 1$, the particles build up near the interface and their final distribution in the dry film may develop gradients, while for $\text{Pe} \ll 1$, the particles diffuse fast enough to mitigate evaporative effects and are expected to be uniformly distributed in the deposited film.\cite{Routh2004}

In the case of a suspension of a bidisperse mixture of particles made from the same material but having different diameters, $d_l$ and $d_s$, the final distribution of particles is determined by two P\'{e}clet numbers, $\text{Pe}_l$ and $\text{Pe}_s$, for the large and small particles, respectively. If the Stokes-Einstein relationship holds, then $\text{Pe}_l/\text{Pe}_s= d_l/d_s>1$. When $\text{Pe}_l > 1 > \text{Pe}_s$, Trueman \textit{et al.} found the so-called ``large-on-top'' stratification,\cite{Trueman2012a, Trueman2012} where the larger (smaller) particles are enriched (depleted) near the receding interface. Recently, Fortini \textit{et al.} discovered the counterintuitive ``small-on-top'' stratification in the regime of $\text{Pe}_l >\text{Pe}_s \gg 1$, i.e., when drying is extremely rapid.\cite{Fortini2016,Fortini2017} Since then, a number of experimental,\cite{Martin-Fabiani2016, Makepeace2017, Liu2018, Cusola2018, Carr2018, Martin-Fabiani2018} theoretical,\cite{Zhou2017, Sear2017,Sear2018} and simulation \cite{Howard2017,Howard2017b,Tang2018,Statt2018,Tatsumi2018} studies have been reported on the stratification phenomena in drying suspensions of polydisperse particles and their physical mechanisms. The idea of diffusiophoresis being responsible for ``small-on-top'' stratification is widely supported.\cite{Fortini2016, Zhou2017, Howard2017, Sear2017, Sear2018, Tang2018} In this picture, when $\text{Pe}_s \gg 1$, the smaller particles congregate near the receding interface during evaporation and their distribution develops a gradient that decays into the drying film. Further, when the volume fraction of the smaller particles, $\phi_s$, is above certain threshold that depends on $\text{Pe}_s$, this gradient generates a diffusiophoretic force that is strong enough to push the larger particles out of the interfacial region. Consequently, the larger particles are depleted near the interface, resulting in ``small-on-top'' stratification.

The key ingredient of the diffusiophoretic model is that the cross-interaction between the large and small particles has asymmetric effects on the phoretic drift of particles and drives the larger ones away from the interfacial region faster than the smaller ones.\cite{Zhou2017, Sear2017} Therefore, the size asymmetry, quantified as $\alpha = d_l/d_s$, is a crucial parameter that controls the outcome of stratification, with larger $\alpha$ favoring ``small-on-top'' stratification. Mart\'{i}n-Fabiani \textit{et al.} studied a system with the smaller particles coated with hydrophilic shells and explored the effect of changing the pH of the initial dispersion.\cite{Martin-Fabiani2016} In a dispersion with low pH, $\alpha$ is large enough to lead to ``small-on-top'' stratification. When the pH is raised, $\alpha$ is reduced as the hydrophilic shells swell substantially, and stratification is suppressed.

The approach of Mart\'{i}n-Fabiani \textit{et al.} can be used for systems where the particle size can be tuned with external stimuli.\cite{Martin-Fabiani2016} However, other possible approaches of controlling stratification for systems with fixed particle sizes have rarely been explored. In a previous work,\cite{Tang2018} we used molecular dynamics (MD) modeling to study drying suspensions of a binary mixture of nanoparticles and found that for fast evaporation rates, the solvent can develop a negative temperature gradient toward the interface because of evaporative cooling effect. This temperature gradient induces thermophoresis, in which the larger particles are pushed more strongly into the interfacial region where the temperature is lower and the solvent density is higher. The competition between thermophoresis generated by evaporative cooling and diffusiophoresis can thus suppress ``small-on-top'' stratification at ultrafast drying rates or even turn the stratification into ``large-on-top''.\cite{Tang2018}  This discovery further indicates that thermophoresis, with a controlled thermal gradient other than the naturally occurring evaporative cooling, may be used to control stratification. In this paper, we employ MD modeling to test this idea in detail and demonstrate that stratification in a drying suspension can be controlled on demand with a temperature gradient imposed on the system, i.e., via controlled thermophoresis.

\section{METHODS}

We performed MD simulations on a suspension of a bidisperse mixture of nanoparticles.\cite{Tang2018} The solvent is modeled explicitly as beads of mass $m$ and interacting with each other via a Lennard-Jones (LJ) potential, $U_\text{LJ}(r) = 4\epsilon \left[ (\sigma/r)^{12} - (\sigma/r)^6 - (\sigma/r_c)^{12} + (\sigma/r_c)^6 \right]$, where $r$ is the center-to-center distance between beads, $\epsilon$ is an energy scale, $\sigma$ is a length scale, and the potential is truncated at $r_c = 3 \sigma$. The nanoparticles are modeled as spheres with a uniform distribution of LJ beads at a mass density $1.0 m/\sigma$.\cite{Everaers2003, IntVeld2008} The large nanoparticles (LNPs) have diameter $d_l = 20 \sigma$ and mass $m_l = 4188.8m$, and the small nanoparticles (SNPs) have diameter $d_s = 5 \sigma$ and mass $m_s = 65.4m$. The size ratio is $\alpha= 4$. The nanoparticle-nanoparticle interactions are given by an integrated form of the LJ potential for two spheres with a Hamaker constant, $A_\text{nn}$, characterizing the interaction strength.\cite{Everaers2003, IntVeld2008} In this study, $A_\text{nn}=39.48 \epsilon$. To ensure that nanoparticles are well dispersed in the initial suspension, the nanoparticle-nanoparticle interactions are rendered purely repulsive by truncating them at $20.574 \sigma$, $13.085\sigma$, and $5.595\sigma$ for the LNP-LNP, LNP-SNP, and SNP-SNP pairs, respectively. The nanoparticle-solvent interactions are described by a similar integrated form of the LJ potential with a Hamaker constant $A_\text{ns} = 100\epsilon$ and a cutoff length $d/2 + 4\sigma$, where $d$ is the nanoparticle diameter.\cite{Cheng2012} The nanoparticle-solvent interactions thus have attractive tails, which make the effective diameter of a nanoparticle larger than its nominal diameter.\cite{Tang2018}. The size ratio is defined here based on the nominal diameters of LNPs and SNPs. If their effective diameters are used, then the size ratio is about 3.4.

The entire system consists of $\sim 7\times10^{6}$ LJ beads, 200 LNPs, and 6400 SNPs. The system is placed in a rectangular simulation cell of dimensions $L_x \times L_y \times L_z$, where $L_x = L_y = 201 \sigma$, and $L_z = 477 \sigma$. The liquid-vapor interface is in the $x$-$y$ plane, in which periodic boundary conditions are imposed. In the initial suspension, the thickness of the liquid film is about $304 \sigma$. The volume fractions of LNPs and SNPs in the initial dispersion are $\phi_l = 0.068$ and $\phi_s = 0.034$, respectively. Along the $z$-axis, all particles are confined in the simulation cell by two walls at $z = 0$ and $z = L_z$. The particle-wall interaction is given by a LJ-like 9-3 potential, $U_W (h) = \epsilon_W \left[ (2/15)(D_W/h)^9 - (D_W/h)^3 - (2/15)(D_W/h_c)^9 +\right.$ $\left. (D_W/h_c)^3 \right]$, where the interaction strength $\epsilon_W = 2.0\epsilon$, $h$ is the distance between the particle center and the wall, and $h_c$ is the cutoff length of the potential. For the solvent beads, $D_W = 1\sigma$ and $h_c = 3\sigma$ ($0.8583\sigma$) at the lower (upper) wall. With these parameters, the liquid solvent completely wets the lower wall while the upper wall is purely repulsive. For the nanoparticles, both walls are repulsive with $D_W = d/2$ and $h_c = 0.8583D_W$, where $d$ is the nanoparticle diameter.

To model evaporation of the solvent, a rectangular box of dimensions $L_x \times L_y \times 20\sigma$ at the top of the simulation cell is designated as a deletion zone and a certain number ($\zeta$) of vapor beads of the solvent in this zone are removed every $\tau$, where $\tau= \sigma (m/\epsilon)^{1/2}$ is the reduced LJ unit of time. In this paper, two evaporation rates $\zeta = 30$ and $\zeta = 5$ are adopted. At these rates, the liquid-vapor interface retreats during evaporation at almost a constant speed, $v_e$. The value of $v_e$ is determined for each evaporating suspension by directly computing the location of the interface as a function of time. The diffusion coefficients of nanoparticles are calculated with direct, independent simulations and the results are $D_l = 3.61 \times 10^{-3} \sigma^2/\tau$ for LNPs and $D_s = 2.11 \times 10^{-2} \sigma^2/\tau$ for SNPs at the initial volume fractions of nanoparticles prior to evaporation (see Supporting Information). The ratio $D_s/D_l = 5.8$ is higher than $\alpha=4$, the value expected from the Stokes-Einstein relation, which may be due to the finite concentrations of nanoparticles.\cite{IntVeld2009} With values of $D_l$, $D_s$, $v_e$, and $H$ determined, the P\'{e}clet numbers for LNPs and SNPs, $\text{Pe}_l$ and $\text{Pe}_s$, are computed for each evaporating system.

The Large-scale Atomic/Molecular Massively Parallel Simulator (LAMMPS)\cite{Plimpton1995} was employed for all simulations reported here. A velocity-Verlet algorithm with a time step $\delta t = 0.01 \tau$ was used to integrate the equation of motion. We have performed tests to confirm that the results reported here remain unchanged if a smaller time step is used. In the thermalized zone(s) specified for each system, a Langevin thermostat with a small damping rate $\Gamma = 0.01 \tau^{-1}$ was used for the solvent beads. We have confirmed that this weak damping is strong enough to ensure a constant temperature in each thermalized liquid zone.

All results are presented below in the LJ units. Here we provide a rough mapping of these units to real ones in Table~\ref{tb:mapping} by mapping the LJ solvent adopted in this paper to a liquid with a critical point similar to water, a typical solvent used in drying experiments.\cite{Schulz2018}. The details of this mapping are provided in the Supporting Information.

\begin{table}[htp]
\centering
\begin{tabular}{|c|c|c|} \hline
Physical Quantity & LJ Unit & SI Value\\ \hline
energy & $\epsilon$ & $7.6\times 10^{-21}$ J \\ \hline
length & $\sigma$ & $0.35\times 10^{-9}$ m \\ \hline
mass & $m$ & $4.5\times 10^{-26}$ kg \\ \hline
time & $\tau$ & $0.85\times 10^{-12}$ s \\ \hline
temperature & $\epsilon/k_\text{B}$ & $550$ K \\ \hline
velocity & $\sigma/\tau$ & $4.1\times 10^2$ m/s \\ \hline
diffusion coefficient & $\sigma^2/\tau$ & $1.4\times 10^{-7}~\text{m}^2$/s  \\ \hline
density & $m/\sigma^3$ & $1.05\times 10^3$ kg/$\text{m}^3$  \\ \hline
viscosity & $m/(\tau \sigma)$ & $1.5\times 10^{-4}$ Pa s\\ \hline
pressure & $\epsilon/\sigma^3$ & $1.8\times 10^2$ MPa\\ \hline
\end{tabular}
\caption{Rough mapping between LJ and real units.}
\label{tb:mapping}
\end{table}

\section{RESULTS AND DISCUSSION}

\begin{figure}[tp]
\centering
\includegraphics[width = 0.4\textwidth]{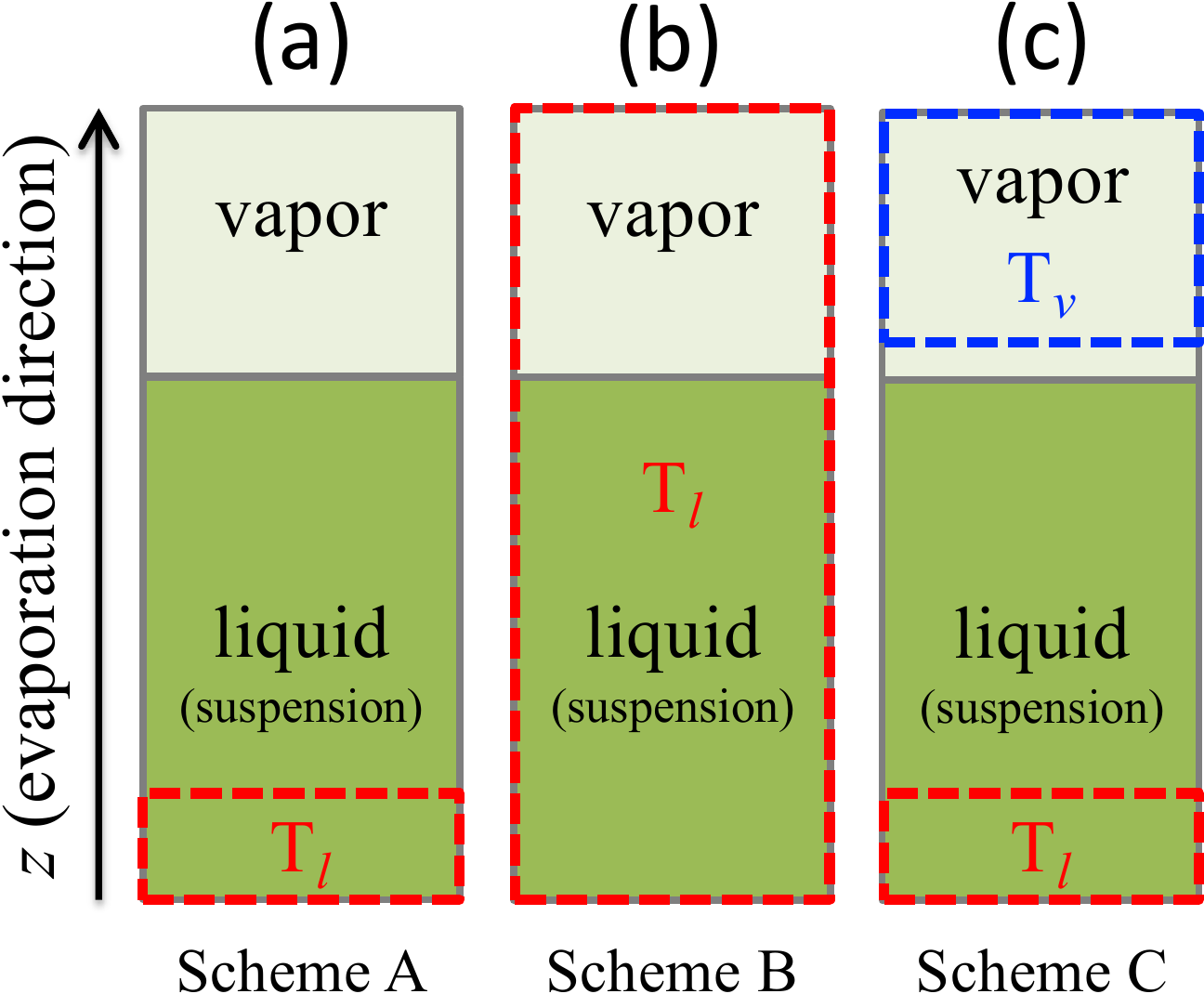}
\caption{Schematics of three types of thermalizations during solvent evaporation: (a) Only a thin layer of the liquid solvent adjacent to the bottom wall is thermalized at $T_l$; (b) All liquid and vapor are thermalized at $T_l$; (c) A thin layer of the liquid solvent adjacent to the bottom wall is thermalized at $T_l$ while the vapor zone at some distance away from the equilibrium liquid-vapor interface is thermalized at $T_v$. We set $T_l=1.0\epsilon/k_\text{B}$ and $T_v$ can be higher or lower than $T_l$ to create a thermal gradient.}
\label{fg:T_scheme}
\end{figure}

Our goal is to demonstrate that a temperature gradient and the associated thermophoretic effect can be used to control stratification in a drying suspension of a polydisperse mixture of nanoparticles. We have previously shown that particles of different sizes have different thermophoretic responses to a thermal gradient.\cite{Tang2018} In our previous work, only a thin layer of the liquid solvent adjacent to the bottom wall is thermalized at $T_l$ during evaporation, as shown in Fig.~\ref{fg:T_scheme}(a). Because of evaporative cooling at the liquid-vapor interface, a negative temperature gradient develops and its magnitude is larger for faster evaporation rates. The negative thermal gradient induces a positive gradient of the solvent density toward the interface, which generates a driving force to transport nanoparticles into the interfacial region.\cite{Piazza2008, Brenner2011PRE} The thermophoretic driving force is stronger for larger particles. The Soret coefficient, $S_T$, can be used to characterize the strength of thermophoretic motion with respect to diffusive motion of particles. We have performed independent simulations of thermophoresis at $A_\text{ns} = 100\epsilon$ and found that for the LNPs, $S_T \sim 0.1~\text{K}^{-1}$, while for the SNPs, their thermophoretic response is extremely weak and $S_T$ is almost 0 (see Supporting Information). As a result, for very fast evaporation relatively more LNPs than SNPs are driven toward the interface in a drying bidisperse suspension.\cite{Tang2018} The thermophoresis caused by evaporative cooling competes with the diffusiophoresis that leads to ``small-on-top'' stratification at fast drying rates, which is why only weak ``small-on-top'' stratification was observed in our previous simulations.\cite{Tang2018} In certain cases the ``small-on-top'' stratification expected by the existing theory \cite{Sear2017} was even converted to ``large-on-top'' in the presence of strong evaporative cooling.\cite{Tang2018}

Based on the physical picture depicted above, it is natural to investigate the effects of a controlled thermal gradient on stratification in a drying suspension. In this paper, we explore this idea by comparing three types of thermalization schemes as sketched in Fig.~\ref{fg:T_scheme}. The Scheme A is the same as in our previous work in which only a $10\sigma$ thick layer of the liquid solvent at the bottom of the suspension is thermalized at $T_l$ [Fig.~\ref{fg:T_scheme}(a)].\cite{Tang2018} Evaporative cooling leads to a negative temperature gradient in the suspension toward the interface. In Scheme B, all solvent beads in the simulation cell are thermalized at $T_l$ [Fig.~\ref{fg:T_scheme}(b)] and thus there are no thermal gradients during evaporation. In Scheme C, in addition to a liquid layer of thickness $10\sigma$ thermalized at $T_l$ near the bottom wall, the vapor beads with $z$-coordinates between $L_z - 150\sigma$ and $L_z$ are coupled to a thermostat with a target temperature $T_v$ [Fig.~\ref{fg:T_scheme}(c)]. In this way, a thermal gradient is generated in the suspension with its direction and magnitude controlled by the difference between $T_v$ and $T_l$, the thickness of the film, and the strength of evaporative cooling (i.e., the evaporation rate). For all systems studied in this paper, $T_l=1.0\epsilon/k_\text{B}$. For Scheme C, $T_v$ is varied from $0.75\epsilon/k_\text{B}$ to $1.2\epsilon/k_\text{B}$.

\begin{table*}[htp]
\centering
\caption{Parameters for all systems studied. Refer to Fig.~\ref{fg:T_scheme} for the thermalization schemes.}
\begin{tabular}{llllll}
\hline
System & $\zeta$  & $v_e\tau/\sigma$         & $\text{Pe}_l$ & $\text{Pe}_s$ & Thermalization Scheme\\ \hline
$T^l_{1.0}\zeta_{30}$     & 30       & 1.13$\times 10^{-3}$   & 95.2         & 16.3         & A         \\
$T^l_{1.0}\zeta_{5}$     & 5        & 2.04$\times 10^{-4}$   & 17.2          & 2.9           & A  \\
$T_{1.0}\zeta_{30}$     & 30       & 1.18$\times 10^{-3}$   & 99.4         & 17.0                 & B    \\
$T_{1.0}\zeta_{5}$     & 5        & 2.11$\times 10^{-4}$   & 17.8          & 3.0           & B     \\
$T^l_{1.0}T^v_{1.2}\zeta_{5}$     & 5        & 2.04$\times 10^{-4}$   & 17.2          & 2.9           & C, $T_v =1.2\epsilon/k_{\rm B}$ \\
$T^l_{1.0}T^v_{1.1}\zeta_{5}$     & 5        & $1.99 \times 10^{-4}$   & 16.8          & 2.9           & C, $T_v =1.1\epsilon/k_{\rm B}$ \\
$T^l_{1.0}T^v_{1.05}\zeta_{5}$     & 5        & $2.04 \times 10^{-4}$   & 17.2          & 2.9         & C, $T_v =1.05\epsilon/k_{\rm B}$ \\
$T^l_{1.0}T^v_{1.0}\zeta_{5}$     & 5        & $2.07 \times 10^{-4}$   & 17.4          & 3.0         & C, $T_v =1.0\epsilon/k_{\rm B}$ \\
$T^l_{1.0}T^v_{0.9}\zeta_{5}$     & 5        & $6.93 \times 10^{-4}$   & 58.4          & 10.0           & C, $T_v =0.9\epsilon/k_{\rm B}$ \\
$T^l_{1.0}T^v_{0.85}\zeta_{5}$     & 5        & $9.90 \times 10^{-4} $   & 83.4         & 14.3           & C, $T_v = 0.85\epsilon/k_{\rm B}$ \\
$T^l_{1.0}T^v_{0.75}\zeta_{5}$     & 5        & $1.03 \times 10^{-3}$   & 86.7         & 14.8           & C, $T_v = 0.75\epsilon/k_{\rm B}$ \\
\hline
\end{tabular}
\label{tb:system}
\end{table*}

For Scheme A, the systems are labeled as $T^l_{1.0}\zeta_{y}$ where the subscript $y$ denotes the value of $\zeta$. For Scheme B, $T_{1.0}\zeta_{y}$ is used to emphasize that the entire system is maintained at $1.0\epsilon/k_\text{B}$ during evaporation. For Scheme C, the systems are labeled as $T^l_{1.0}T^v_{x}\zeta_{y}$, where $x$ indicates the value of $T_v$. All systems studied are listed in Table~\ref{tb:system}. $T^l_{1.0}T^v_{1.1}\zeta_{5}$, $T^l_{1.0}T^v_{1.05}\zeta_{5}$, and $T^l_{1.0}T^v_{1.0}\zeta_{5}$ have results in line with $T^l_{1.0}T^v_{1.2}\zeta_{5}$. We also studied systems with $\zeta = 5$ and $T_v < T_l$, which show negative thermal gradients in the suspension and thermophoresis similar to those in $T^l_{1.0}\zeta_{30}$ and $T^l_{1.0}\zeta_{5}$ where evaporative cooling occurs. However, we observed condensation of droplets in the vapor phase if $T_v$ is made lower than the temperature at the liquid-vapor interface in Scheme A with the same $\zeta$. Despite this unwanted effect, cooling the vapor at a temperature lower than that of the suspension could be one experimental approach to apply a negative thermal gradient for systems that evaporate slowly or for which the effect of evaporative cooling is not as strong as that for the model LJ liquid employed in our simulations. The last six systems in Table~\ref{tb:system} with $T_v$ varying from $0.75\epsilon/k_\text{B}$ to $1.1\epsilon/k_\text{B}$ are included in the Supporting Information. In the main text we focus on the first five systems in Table~\ref{tb:system}.

\begin{figure}[tp]
\centering
\includegraphics[width = 0.45\textwidth]{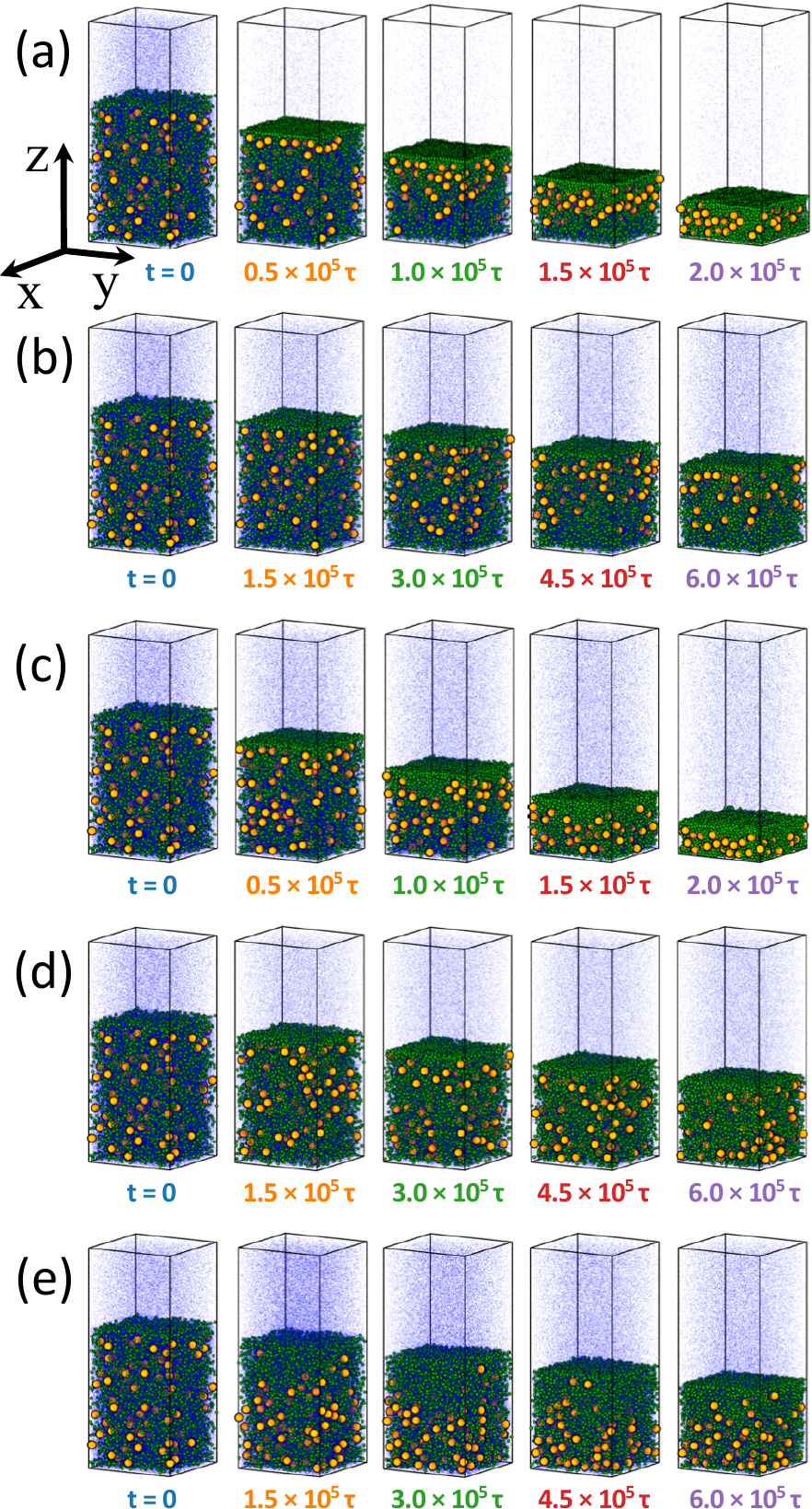}
\caption{Snapshots during solvent evaporation for (a) $T^l_{1.0}\zeta_{30}$, (b) $T^l_{1.0}\zeta_{5}$, (c) $T_{1.0}\zeta_{30}$, (4) $T_{1.0}\zeta_{5}$, and (5) $T^l_{1.0}T^v_{1.2}\zeta_{5}$. Elapsed time since the initiation of evaporation at $t=0$ is listed under each snapshot. Temperature and density profiles of the five systems are shown in Fig.~\ref{fg:density}. Color code: SNPs (green), LNPs (orange), and solvent (blue). Only 5\% of the solvent beads are visualized to improve clarity.}
\label{fg:snaps}
\end{figure}

Snapshots of the first five nanoparticle suspensions in Table~\ref{tb:system} during solvent evaporation are shown in Fig.~\ref{fg:snaps}. For $T^l_{1.0}\zeta_{30}$ and $T^l_{1.0}\zeta_{5}$ [Figs.~\ref{fg:snaps}(a) and (b)], the evaporative cooling of the liquid-vapor interface leads to a negative thermal gradient along the normal direction toward the interface. Although for both systems ``small-on-top'' stratification is expected by the model of Zhou \textit{et al.} since $\text{Pe}_l \gg \text{Pe}_s > 1$,\cite{Zhou2017} the thermophoresis associated with the negative temperature gradient works against diffusiophoresis and transports more LNPs into the interfacial region. As a result, the two systems exhibit ``large-on-top'' stratification.

When all solvent beads in the simulation cell are thermalized during evaporation, the temperature in the entire system is constant and no thermal gradients are produced. Thermophoresis is thus suppressed and only diffusiophoresis remains. The expected outcome is ``small-on-top'' stratification for $\text{Pe}_l \gg \text{Pe}_s > 1$. The results from $T_{1.0}\zeta_{30}$ and $T_{1.0}\zeta_{5}$ confirm this prediction, as shown in Figs.~\ref{fg:snaps}(c) and (d). For example, comparing the last snapshot for $T^l_{1.0}\zeta_{5}$ (the second row of Fig.~\ref{fg:snaps}) and that for $T_{1.0}\zeta_{5}$ (the fourth row of Fig.~\ref{fg:snaps}), the transition from ``large-on-top'' to ``small-on-top'' is visible after the thermal gradients and the associated thermophoresis are inhibited, especially from the distribution of LNPs in the drying films. This transition is verified quantitatively by an order parameter of stratification, which is discussed below (see Fig.~\ref{fg:op}).

The last row of Fig.~\ref{fg:snaps} shows the snapshots for $T^l_{1.0}T^v_{1.2}\zeta_{5}$. In this system, the vapor about $23\sigma$ above the initial liquid-vapor interface prior to evaporation are thermalized at $T_v = 1.2\epsilon/k_\text{B} > T_l$ during evaporation. Consequently, there is a positive temperature gradient in the liquid solvent along the film's normal direction toward the interface. The solvent density develops a negative gradient and the accompanied thermophoresis drives LNPs toward the substrate. As a result, thermophoretic and diffusiophoretic effects are in synergy and strong ``small-on-top'' stratification is generated, which is apparent in Fig.~\ref{fg:snaps}(e) where the LNPs are enriched near the substrate during drying.

\begin{figure*}[tp]
\centering
\includegraphics[width = 1 \textwidth]{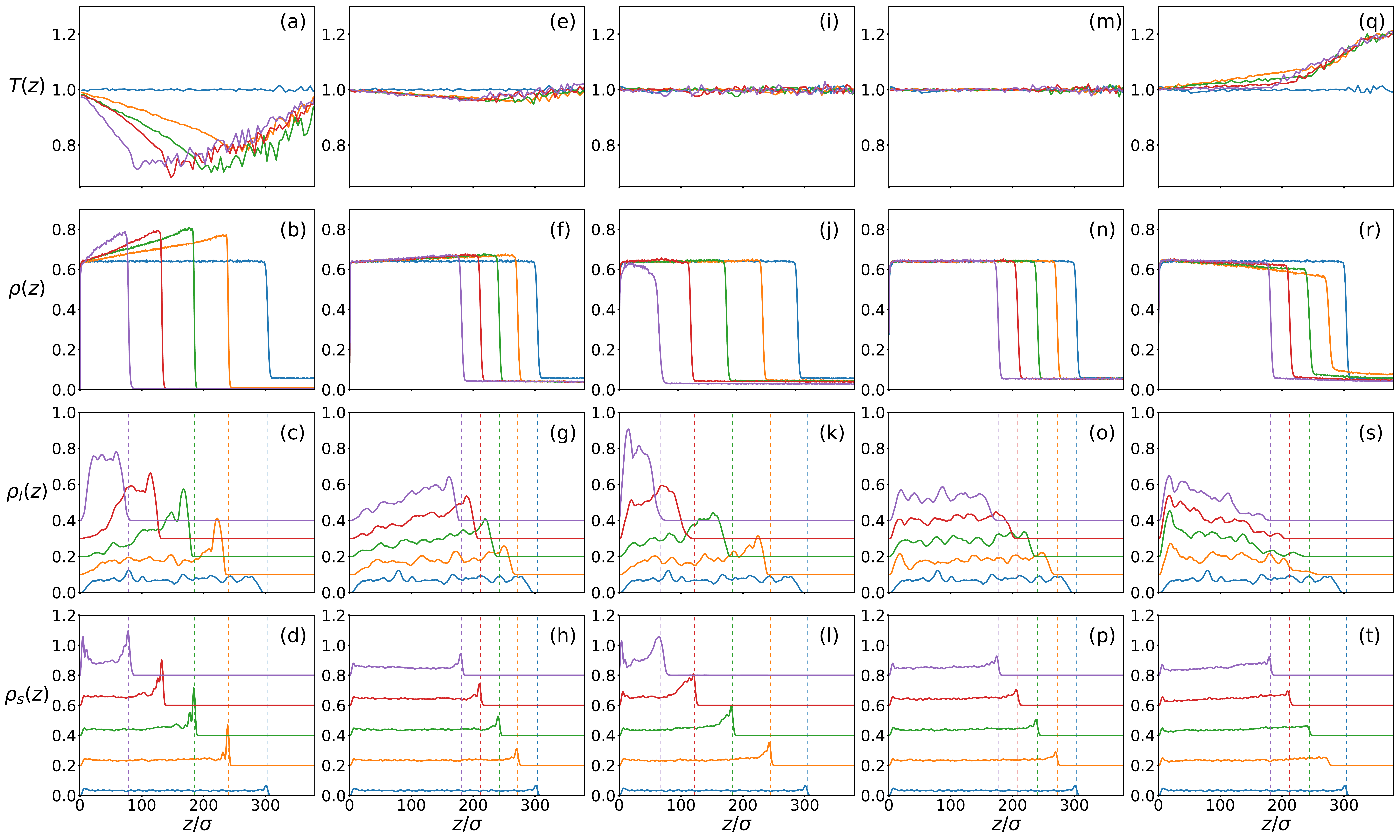}
\caption{Temperature profiles (top row) and density profiles for the solvent (second row), LNPs (third row), and SNPs (bottom row) for $T^l_{1.0}\zeta_{30}$ (a-d), $T^l_{1.0}\zeta_{5}$ (e-h), $T_{1.0}\zeta_{30}$ (i-l), $T_{1.0}\zeta_{5}$ (m-p), and $T^l_{1.0}T^v_{1.2}\zeta_{5}$ (q-t), respectively. The curves follow the same order as the snapshots shown Fig.~\ref{fg:snaps}. The vertical dashed lines indicate the location of the liquid-vapor interface. For clarity, the density profiles for LNPs (SNPs) are shifted upward by $0.1m/\sigma^3$ ($0.2m/\sigma^3$) successively.}
\label{fg:density}
\end{figure*}

\begin{figure}[tp]
\centering
\includegraphics[width = 0.45\textwidth]{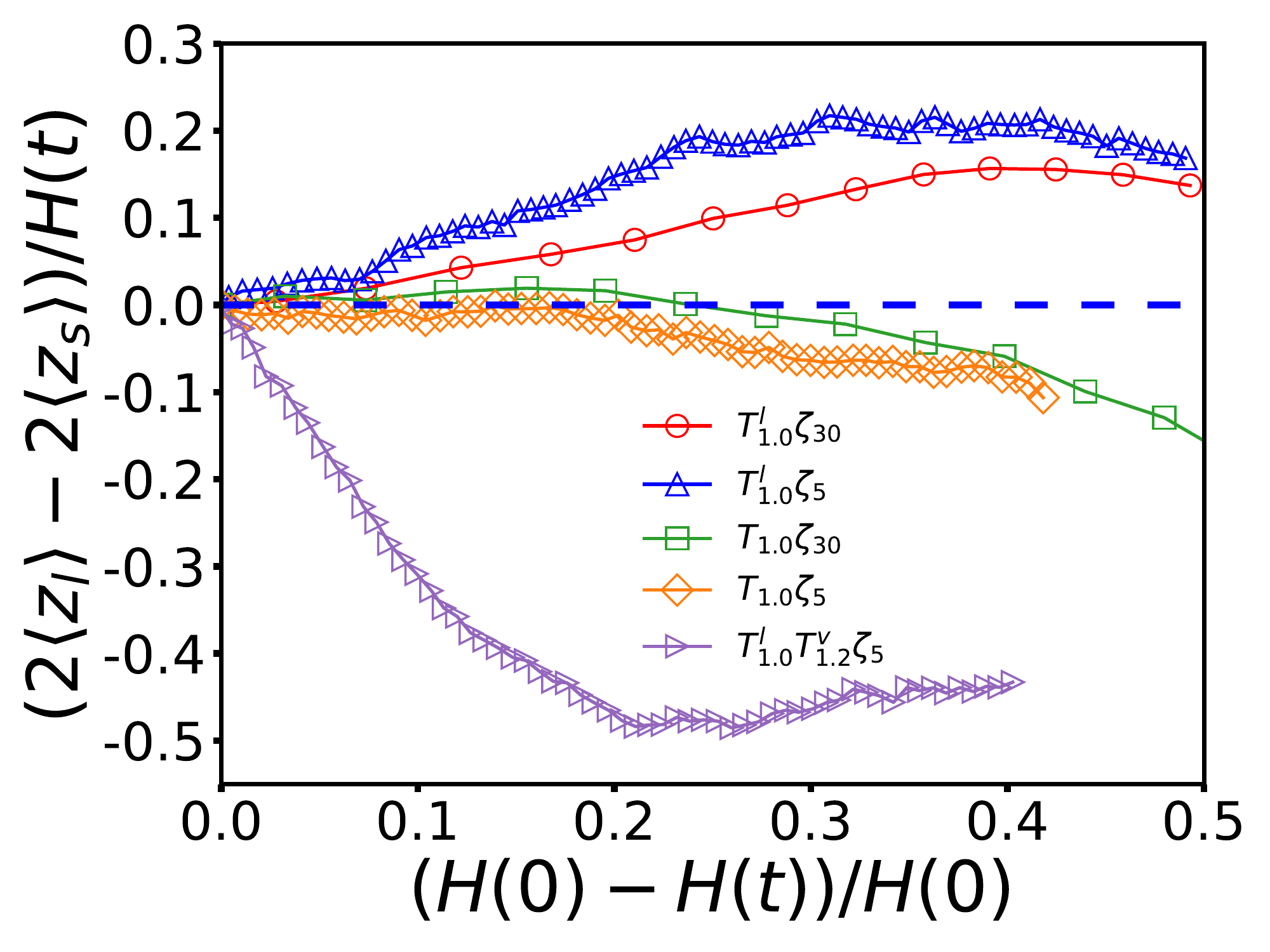}
\caption{Mean separation between LNPs and SNPs normalized by $H(t)/2$, \textit{vs} extent of drying, $(H(0)-H(t))/H(0)$, for $T^l_{1.0}\zeta_{30}$ (red circle), $T^l_{1.0}\zeta_{5}$ (blue upward triangle), $T_{1.0}\zeta_{30}$ (green square), $T_{1.0}\zeta_{5}$ (yellow diamond), and $T^l_{1.0}T^v_{1.2}\zeta_{5}$ (purple right-pointing triangle).}
\label{fg:op}
\end{figure}

To understand quantitatively the stratification phenomena in drying particle suspensions, we plot the temperature and density profiles in Fig.~\ref{fg:density}. The local temperature $T(z)$ at height $z$ is computed from the average kinetic energy of the solvent beads in the spatial bin $[z-2.5\sigma,z+2.5\sigma]$.\cite{Cheng2011} The temperature profiles in the top row of Fig.~\ref{fg:density} clearly show the negative thermal gradients induced by evaporative cooling for $T^l_{1.0}\zeta_{30}$ and $T^l_{1.0}\zeta_{5}$, with the effect stronger at larger evaporation rates. $T_{1.0}\zeta_{30}$ and $T_{1.0}\zeta_{5}$ do not exhibit thermal gradients as all solvent is thermalized at $T_l$, as shown in Figs.~\ref{fg:density}(i) and (m). $T^l_{1.0}T^v_{1.2}\zeta_{5}$ with $T_v > T_l$ exhibits an externally imposed positive thermal gradient [Fig.~\ref{fg:density}(q)].

The local density of solvent or nanoparticles is computed as $\rho_{i}(z) = n_{i} (z) m_i / (L_x L_y\sigma)$, where $n_{i} (z)$ represents the number of particles in the spatial bin $[z - 0.5\sigma, z + 0.5\sigma]$ and $m_i$ is the particle mass. A nanoparticles straddling several bins is partitioned based on its partial volume in each bin. When computing the solvent density, the volume occupied by the nanoparticles is subtracted. The second row of Fig.~\ref{fg:density} shows the solvent density as a function of height and the profiles exhibit gradients in accordance with the thermal gradients. Particularly, a positive (negative) thermal gradient generates a negative (positive) density gradient for the solvent and the stronger the thermal gradient, the stronger the density gradient. This correlation results from the fact that local thermal equilibrium is always maintained even at the fastest evaporation rates adopted in our simulations.\cite{Cheng2011} The density profile of the solvent affects the receding speed, $v_e$, of the liquid-vapor interface. The data in Table~\ref{tb:system} show that at the same $\zeta$, the value of $v_e$ is slightly lower for $T^l_{1.0}\zeta_{y}$ under Scheme A than for $T_{1.0}\zeta_{y}$ under Scheme B. For $T^l_{1.0}\zeta_{y}$, the evaporative cooling causes a positive gradient of the solvent density. The average solvent density is thus higher for $T^l_{1.0}\zeta_{y}$ than for $T_{1.0}\zeta_{y}$, as shown in Figs.~\ref{fg:density}(b), (f), (j), and (n). As a result, the liquid-vapor interface recedes more slowly in $T^l_{1.0}\zeta_{y}$ than in $T_{1.0}\zeta_{y}$ at the same $\zeta$.

The density profiles for LNPs and SNPs are shown in the bottom two rows of Fig.~\ref{fg:density}, respectively. These profiles demonstrate the phoretic response of the nanoparticles to the thermal gradients (equivalently, the density gradients of the solvent induced by the thermal gradients) as well as the effects of the evaporation rate. For all simulations discussed here, the evaporation rates are high enough such that $\text{Pe}_l \gg \text{Pe}_s > 1$. The corresponding fast receding liquid-vapor interface tends to trap both LNPs and SNPs just below the interface. If no other factors are at play, this effect combined with a large enough $\phi_s$ is expected to yield ``small-on-top'' stratification via the diffusiophoresis mechanism as suggested by Sear and collaborators\cite{Fortini2016,Sear2017} and Zhou \textit{et al.}.\cite{Zhou2017} This scenario is indeed the case for $T_{1.0}\zeta_{30}$ and $T_{1.0}\zeta_{5}$, as shown in the third and fourth columns of Fig.~\ref{fg:density} where there are no thermal gradients. The diffusiophoresis model also implies that the degree of ``small-on-top'' stratification is enhanced when the evaporation rate is increased.\cite{Fortini2016, Zhou2017} However, as shown later, $T_{1.0}\zeta_{5}$ actually exhibits stronger ``small-on-top'' stratification than $T_{1.0}\zeta_{30}$, even though the evaporation rate is increased six fold in the latter system. This discrepancy may be partially due to the small thickness of the suspension film studied in our simulations, which is limited by the available computational resources. The effect of film thickness on stratification is explored in a separate study.\cite{Tang2018arXiv_compare} Another reason may be that when the evaporation rate is increased, the drying time is shortened and there is less time for LNPs to diffuse out of the interfacial region via diffusiophoresis. As a result, ``small-on-top'' stratification is weakened when the evaporation rate is enhanced beyond certain threshold. This trend indicates that ``small-on-top'' stratification is most enhanced at some $\text{Pe}_l$ and is diminished if $\text{Pe}_l$ is increased further, which is consistent with two recent reports.\cite{Tatsumi2018, Tang2018arXiv_compare}

When only a thin layer of solvent beads at the bottom wall is thermalized, the temperature in the vicinity of the liquid-vapor interface decreases because of evaporative cooling effect. The resulting enhancement of the solvent density at the interface leads to thermophoretic drift of nanoparticles with the effect more significant for larger particles. This physical picture explains the observations for $T^l_{1.0}\zeta_{30}$ and $T^l_{1.0}\zeta_{5}$. In these two systems, the SNPs are found to accumulate at the surface of the evaporating suspension as $\text{Pe}_l \gg \text{Pe}_s > 1$ [Figs.~\ref{fg:density}(d) and (h)]. However, a significant accumulation of LNPs is found just below the enriched surface layer of SNPs, as shown in Figs.~\ref{fg:density}(c) and (g). The net outcome is actually ``large-on-top'' stratification, which will be confirmed later with an order parameter quantifying stratification (see Fig.~\ref{fg:op}). Furthermore, the degree of ``large-on-top'' stratification is stronger for $T^l_{1.0}\zeta_{5}$ than for $T^l_{1.0}\zeta_{30}$, indicating a delicate competition between diffusiophoresis and thermophoresis. The lower evaporation rate in $T^l_{1.0}\zeta_{5}$ suppresses both processes but it appears that diffusiophoresis favoring ``small-on-top'' is mitigated slightly more, creating stronger ``large-on-top'' for $T^l_{1.0}\zeta_{5}$.

In our previous work,\cite{Tang2018} we obtained a state diagram of stratification with systems all thermalized with Scheme A (i.e., a thin layer of liquid solvent contacting the substrate is thermalized at $T_l = 1.0\epsilon/k_\text{B}$) and only observed weak ``small-on-top'' stratification at values of $\text{Pe}_s$ and $\phi_s$ far exceeding the critical values predicted by the diffusiophoretic model of Zhou \textit{et al.}.\cite{Zhou2017} The presence of thermophoresis at fast evaporation rates may help understand the discrepancy between the simulations and the theory.\cite{Tang2018} Indeed, when thermophoresis is suppressed, systems that are driven into the ``large-on-top'' regime by thermophoresis can be turned into (usually weak) ``small-on-top''. Examples are the transition from $T^l_{1.0}\zeta_{30}$ to $T_{1.0}\zeta_{30}$ and that from $T^l_{1.0}\zeta_{5}$ to $T_{1.0}\zeta_{5}$.

To achieve strong ``small-on-top'' stratification, a natural idea is to enable thermophoresis that works in conjunction with diffusiophoresis. This cooperation requires a thermal gradient during evaporation that is opposite to the one induced by evaporative cooling. To realize this, we thermalize the vapor zone from $L_z - 150\sigma$ to $L_z$ at a temperature $T_v > T_l$. The data in the fifth column of Fig.~\ref{fg:density} are for $T^l_{1.0}T^v_{1.2}\zeta_{5}$ where $T_v = 1.2\epsilon/k_\text{B}$. A positive thermal gradient and a negative density gradient of the solvent can be seen clearly in Figs.~\ref{fg:density}(q) and (r), respectively. Since the gradients are reversed, the LNPs are now driven toward the substrate via thermophoresis [Fig.~\ref{fg:density}(s)] while the SNPs are much less affected [Fig.~\ref{fg:density}(t)]. The final result is strong ``small-on-top'' stratification where the LNPs are accumulated near the substrate and depleted in the interfacial region while the SNPs exhibit a positive density gradient (i.e., accumulation) from the bulk of the film to the receding interface as $\text{Pe}_s > 1$.

It is expected that for systems thermalized with Scheme C and $T_v < T_l$, a negative thermal gradient develops in the liquid solvent, similar to the evaporative cooling case in Scheme A. Consequently, systems under Scheme C with $T_v < T_l$ could display ``large-on-top'' stratification as long as the thermal gradient is large enough. These cases are in fact observed and discussed in detail in the Supporting Information, where some complications are noted related to droplet condensation in a vapor that is thermalized at low temperatures. Even for $T_v \gtrsim T_l$, the thermal gradient in the drying suspension can still be negative if evaporative cooling is strong enough. This is the case for $T^l_{1.0}T^v_{1.05}\zeta_{5}$ and $T^l_{1.0}T^v_{1.0}\zeta_{5}$ (see Supporting Information).

To quantify stratification, we define an order parameter using the full density profiles of nanoparticles.\cite{Tang2018} The mean heights of LNPs and SNPs are computed as $\langle z_i \rangle= \frac{1}{N_i} \sum\limits_{n=1}^{N_i} z_{in}$ with $i \in \{l, s\}$. The order parameter of stratification is then computed as $ (2\langle z_l \rangle -2 \langle z_s\rangle)/H(t) $, i.e., the mean separation between LNPs and SNPs normalized by $H(t)/2$, where $H(t)$ is the instantaneous thickness of the suspension. In the equilibrium suspension prior to evaporation, both $ \langle z_l \rangle $ and $ \langle z_s \rangle $ are very close to $H(0)/2$, where $H(0)$ is the initial film thickness. During evaporation, $ \langle z_l \rangle - \langle z_s\rangle < 0$ indicates ``small-on-top'' stratification while  $ \langle z_l \rangle - \langle z_s\rangle > 0$ signifies ``large-on-top''.

In Fig.~\ref{fg:op} the order parameter of stratification is plotted against the extent of drying, quantified as $(H(0)-H(t))/H(0)$, for the first five systems listed in Table~\ref{tb:system}. It is clear that $T_{1.0}\zeta_{30}$ and $T_{1.0}\zeta_{5}$ exhibit ``small-on-top'' stratification since diffusiophoresis dominates while thermal gradients and thermophoresis are absent. The extent of stratification is slightly stronger for $T_{1.0}\zeta_{5}$, though it dries more slowly.  ``Large-on-top'' is observed for $T^l_{1.0}\zeta_{30}$ and $T^l_{1.0}\zeta_{5}$ and is again stronger for $T^l_{1.0}\zeta_{5}$ that has a smaller evaporation rate. Although thermophoresis is much weaker for $T^l_{1.0}\zeta_{5}$ because of the reduced evaporation rate, diffusiophoresis favoring ``small-on-top'' is suppressed even more when evaporation is slowed down and the delicate interplay of the two phoretic processes leads to stronger ``large-on-top'' stratification for $T^l_{1.0}\zeta_{5}$.

A dramatic ``small-on-top'' state is clearly demonstrated in Fig.~\ref{fg:op} for $T^l_{1.0}T^v_{1.2}\zeta_{5}$. Note that in the equilibrium suspension, $\phi_l = 2\phi_s$. If in the final dry film all SNPs were on top of all LNPs (i.e., complete stratification) but each group is uniformly distributed in its own region, then $ \langle z_l \rangle = H(t)/3$ and $ \langle z_s \rangle = 5H(t)/6$, yielding $ (2\langle z_l \rangle -2 \langle z_s\rangle)/H(t) = -1 $. As shown in Fig.~\ref{fg:op}, the order parameter of stratification reaches a minimal value around $-0.5$ for $T^l_{1.0}T^v_{1.2}\zeta_{5}$, indicating that the vertical distribution of the binary mixture of nanoparticles is substantially segregated in the drying film with SNPs on top of LNPs. This outcome is visually apparent in Fig.~\ref{fg:snaps}(e) as well. 

The stratification order parameter used here and in Ref.~[\onlinecite{Tang2018}] is based on the average position of nanoparticles, which is the first moment of their density profile in the entire drying film. This order parameter describes the systems studied here well and the identification of a stratified state is consistent with the classification based on the overall trend of the density profile of nanoparticles in the bulk of the drying film. Namely, ``small-on-top'' stratification generally corresponds to a negative gradient of the density profile of LNPs and a positive or nearly zero gradient of the density profile of SNPs from the bottom of the film to the top, while ``large-on-top'' is the other way around. However, this order parameter may not be applicable to oscillating density profiles or systems with only local stratification. In these more complicated situations, some other characteristics of the nanoparticle distribution including the higher moments of the density profile may be necessary to classify stratification. For all of our simulations, there is always a layer enriched with SNPs at the top of the drying film because $\text{Pe}_s > 1$. However, it is misleading to call all these systems ``small-on-top''. Instead, information on the nanoparticle distribution in the film below this SNP-rich surface layer should be taken into account as well. The order parameter used here fulfills this goal and yields a more consistent classification scheme for the outcome of stratification.

Evaporative cooling is a natural effect in a fast drying liquid. If a particle suspension is placed on a substrate that is kept at a constant temperature and the suspension undergoes very fast solvent evaporation, then a temperature lower than that of the substrate is expected at the evaporating interface, resulting in a negative thermal gradient in the suspension. $T^l_{1.0}\zeta_{30}$ and $T^l_{1.0}\zeta_{5}$ studied here are set up to mimic such situations. However, it is challenging to maintain a constant temperature or induce a positive thermal gradient along the normal direction toward the interface in a drying suspension, especially when the evaporation rate is high. One possible approach is to dissolve a gas (e.g., $\text{N}_2$, Ar, He, or $\text{CO}_2$) into the solvent (e.g., water). Beaglehole showed that heating a water film with a dissolved gas from above or below produces very different temperature distributions within the liquid.\cite{Beaglehole1987, Kuz1989} When heated from below, a fairly uniform temperature is found throughout the liquid. However, when the liquid is heated from above, a temperature gradient develops in it with the temperature higher at the liquid-vapor interface. Then it may be possible to study the effect of solvent evaporation on the particle distribution in a drying film under isothermal conditions and positive thermal gradients, similar to Scheme B and C.

In most experiments, films are much thicker than those studied here with MD and evaporation rates are much lower by a factor of $10^4$ to $10^5$ for drying at room temperature, about $45\%$ of the critical temperature of water. In these systems, evaporative cooling is negligible and heat transfer is fast enough to make temperature uniform throughout a drying film.\cite{Price2000, Krantz2010} To mimic this situation, Scheme B is used to maintain an isothermal drying film by coupling all solvent beads including vapor to a weak Langevin thermostat with a small damping rate, $\Gamma = 0.01\tau^{-1}$. To address whether hydrodynamic interactions are screened in Langevin dynamics,\cite{Dunweg1993} we run an additional simulation for $T_{1.0}\zeta_{30}$ with the Langevin thermostat replaced by a pairwise thermostat based on dissipative particle dynamics (DPD) with a weak friction coefficient $\gamma = 0.05m/\tau$.\cite{Tsige2004b} With the DPD thermostat, local momentum conservation is preserved throughout the simulation box and hydrodynamic interactions are expected to be fully captured. The results with the DPD thermostat are very close to those discussed here with the Langevin thermostat. These results are included in the Supporting Information. Under Scheme A and C, local momentum conservation is fulfilled away from the thermalized zones. All tests indicate that the Langevin thermostat adopted here is weak enough such that the viscosity of the LJ liquid is only weakly altered and the screening effect on hydrodynamic interactions is negligible.

In all simulations discussed thus far, the temperature of the thermalized liquid zone is always $1.0\epsilon/k_\text{B}$. The highest temperature used for the thermalized vapor zone is $1.2\epsilon/k_\text{B}$, which is close to the critical temperature, $T_c$, of the LJ solvent with $r_c = 3.0\sigma$. Furthermore, all simulations start with systems in which the nanoparticles are uniformly dispersed and then the evaporation process and the thermal gradient are imposed simultaneously. With this approach the interplay between diffusiophoresis and thermophoresis is investigated. To ensure that the physical mechanism of controlling stratification via thermophoresis is applicable to systems with both liquid and vapor temperatures way below $T_c$, we run an additional simulation for $RT^l_{0.9}T^v_{1.0}\zeta_{5}$, i.e., with the bottom layer of the solvent adjacent to the lower wall thermalized at $0.9\epsilon/k_\text{B}$ while the vapor zone above the liquid-vapor interface thermalized at $1.0\epsilon/k_\text{B}$. For $RT^l_{0.9}T^v_{1.0}\zeta_{5}$, the system is first relaxed under the imposed thermal gradient, which causes the LNPs to drift toward the lower wall via thermophoresis. The SNPs are still uniformly dispersed in the film as they are almost irresponsive to the thermal gradient. Then the solvent is evaporated from the relaxed system. The results for $RT^l_{0.9}T^v_{1.0}\zeta_{5}$ are discussed in detail in the Supporting Information and fully consistent with the idea that thermophoresis from a positive thermal gradient from the bulk of a film to its drying front works in synergy with diffusiophoresis to enhance ``small-on-top'', while a negative thermal gradient works against diffusiophoresis to suppress ``small-on-top'' and promote ``large-on-top''.

\section{CONCLUSIONS}

In this paper we focus on how stratification can be controlled in a drying suspension of a bidisperse mixture of nanoparticles via MD simulations with an explicit solvent model. We demonstrate that a thermal gradient and the induced thermophoresis can be used to alter stratification from ``large-on-top'' all the way to strong ``small-on-top''. This strategy is based on the observation that particles of different sizes in a suspension have different responses to a thermal gradient. In particular, larger particles experience a larger driving force that transports them into cooler regions where the solvent density is higher. For $A_\text{ns} = 100\epsilon$ adopted here, the smaller nanoparticles show little or even no response to a thermal gradient. When a suspension undergoes fast drying and only a thin layer of the solvent adjacent to the substrate is thermalized at $T_l$, mimicking an experimental situation where the substrate supporting the suspension is maintained at a constant temperature during solvent evaporation, a negative temperature gradient develops in the suspension because of the evaporative cooling effect that makes the temperature at the evaporating interface to drop below $T_l$. A larger fraction of the larger nanoparticles are driven into the interfacial region via the thermophoresis induced by this thermal gradient. As a result, the fast drying suspensions display ``large-on-top'' stratification instead of ``small-on-top'' expected by the diffusiophoresis model in which the suspension is assumed to be isothermal during evaporation.\cite{Fortini2016,Zhou2017,Sear2017} 

Interestingly, when the entire suspension is maintained at $T_l$ during drying by thermalizing all solvent beads in the simulation cell, they do exhibit ``small-on-top'' stratification at fast evaporation rates, consistent with the prediction of the diffusiophoresis model.\cite{Fortini2016,Zhou2017,Sear2017} However, the degree of stratification is found to be weak, probably due to the fact that $\phi_s$ is small and the liquid film is thin for the simulations reported here. When a positive thermal gradient is induced in the suspension by thermalizing the vapor at a temperature sufficiently higher than $T_l$, all larger nanoparticles are propelled toward the substrate. In this case, the synergy between thermophoresis and diffusiophoresis is underlying the observation of strong ``small-on-top'' stratification. Our results thus reveal a potentially useful strategy of controlling stratification via a regulated thermal gradient in a drying suspension of polydisperse particles.

The film thickness in our simulations prior to evaporation is about $300\sigma \sim 105$ nm. For a temperature difference $0.1\epsilon/k_\text{B}$ across the film, the magnitude of the thermal gradient is about 0.5 K/nm if we take $\epsilon/k_\text{B}\sim 550$ K as in Table~\ref{tb:mapping}. This thermal gradient is several orders of magnitude larger than a typical experimental value. However, the suspensions simulated here are at temperatures not far from the critical temperature of the solvent, which allows the evaporation process of the solvent to be fast enough that can be modeled with MD. As a result, the evaporation rates in the MD simulations are also much higher than those in experiments. Nevertheless, as already discussed in Ref.~[\onlinecite{Tang2018}], such high evaporation rates are needed to drive a sub-micron thin film suspension of nanoparticles into the ``small-on-stop'' regime (i.e., $\text{Pe}_s > 1$), demonstrated \textit{in silico} with our simulations. It is an interesting challenge if such a scenario can be realized experimentally, for example, by bringing the suspension close to the critical point of its solvent, as it may be relevant to the fabrication of multilayered thin film coatings. Because of high evaporation rates and the resulting strong evaporative cooling, large thermal gradients are needed in order to control stratification in drying thin films. For films with micrometer to millimeter thickness as in many experiments,\cite{Schulz2018} evaporation rates and thermal gradients smaller by a factor of $10^4$ to $10^5$ than those employed in MD simulations, i.e., those within the typical experimental range, will be sufficient to drive the systems into the same physical regime where thermophoresis is comparable to diffusiophoresis. In this sense, the results obtained here from MD simulations with thin films are \textit{scalable} to much thicker films studied in experiments.

\section*{Acknowledgement}
Acknowledgment is made to the Donors of the American Chemical Society Petroleum Research Fund (PRF \#56103-DNI6) for support of this research. This research used resources of the National Energy Research Scientific Computing Center (NERSC), a U.S. Department of Energy Office of Science User Facility operated under Contract No. DE-AC02-05CH11231. These resources were obtained through the Advanced Scientific Computing Research (ASCR) Leadership Computing Challenge (ALCC). This work was performed, in part, at the Center for Integrated Nanotechnologies, an Office of Science User Facility operated for the U.S. Department of Energy Office of Science. Sandia National Laboratories is a multimission laboratory managed and operated by National Technology and Engineering Solutions of Sandia, LLC., a wholly owned subsidiary of Honeywell International, Inc., for the U.S. Department of Energy's National Nuclear Security Administration under contract DE-NA0003525. This paper describes objective technical results and analysis. Any subjective views or opinions that might be expressed in the paper do not necessarily represent the views of the U.S. Department of Energy or the United States Government.


\clearpage
\newpage
\onecolumngrid
\renewcommand{\thefigure}{S\arabic{figure}}
\setcounter{figure}{0}    
\renewcommand{\thepage}{SI-\arabic{page}}
\setcounter{page}{1}    
\begin{center}
{\bf SUPPORTING INFORMATION}
\end{center}

\noindent {\bf S1. Diffusion Coefficients of Nanoparticles:}

\begin{figure}[bp]
\includegraphics[width = 0.75\textwidth]{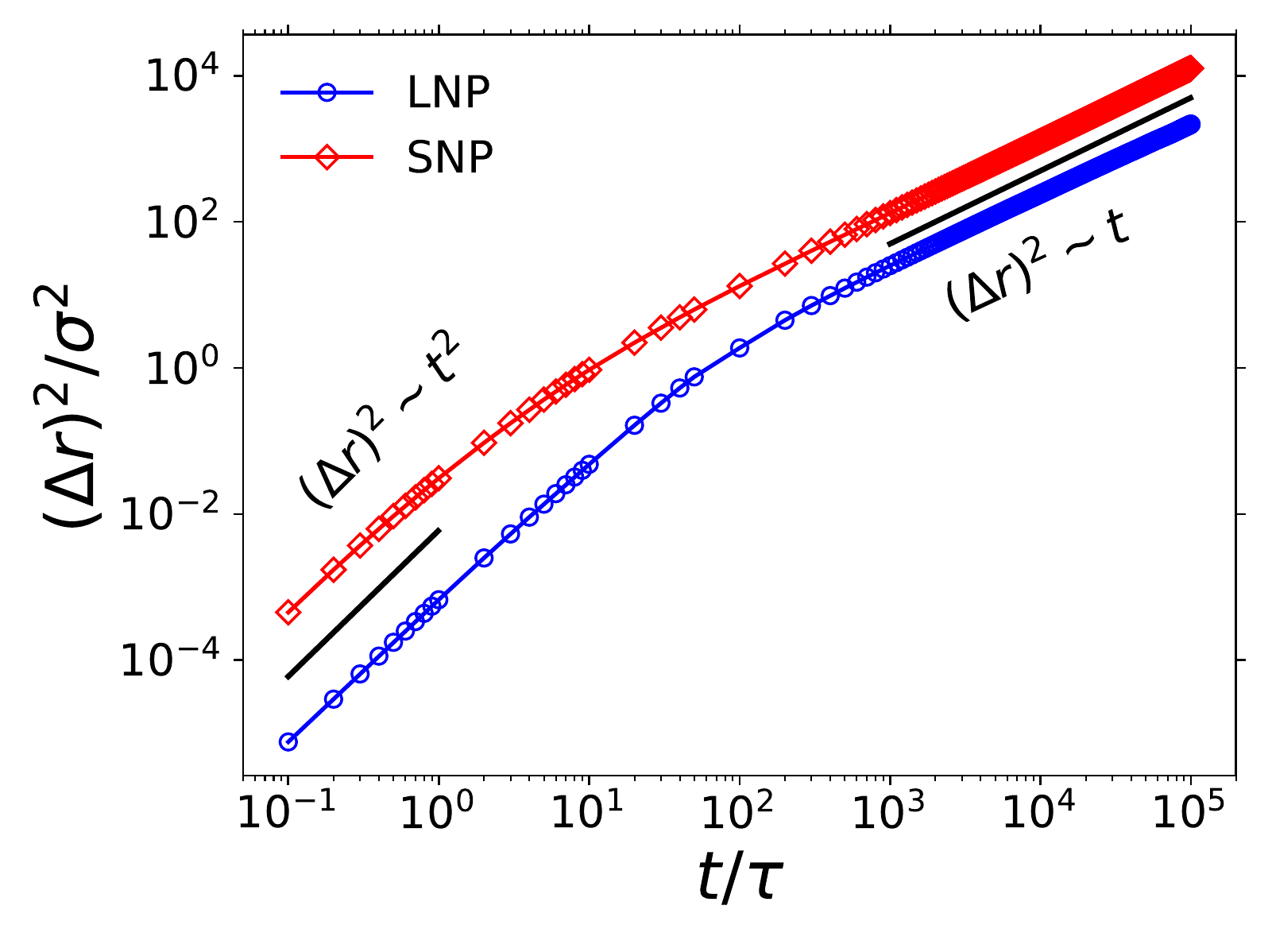}
\caption{Mean square displacement \textit{vs} time for LNPs (blue circles) and SNPs (red diamonds).}
\label{fg:diffusion}
\end{figure}

\noindent The diffusion coefficients of the large nanoparticles (LNPs) and small nanoparticles (SNPs) were determined by an independent simulation. A suspension of LNPs and SNPs with the volume fractions ($\phi_l = 0.060$ and $\phi_s = 0.033$) close to those in the initial suspension discussed in the main text was prepared. The suspension has a cubic shape with edge length $202.6\sigma$. Periodic boundary conditions were used in all three directions. The mean square displacements of both LNPs and SNPs as a function of time are shown in Fig.~\ref{fg:diffusion}. The data show a clear transition from ballistic regime at short times to diffusive regime at long times. The diffusion coefficients are $D_l = 3.61 \times 10^{-3} \sigma^2/\tau$ for LNPs and $D_s = 2.11 \times 10^{-2} \sigma^2/\tau$ for SNPs. A simulation for a cubic simulation cell with edge length $101.3\sigma$ gives similar results.

\noindent {\bf S2. Thermophoresis of Nanoparticles:}

\begin{figure}[tp]
\includegraphics[width = 0.5\textwidth]{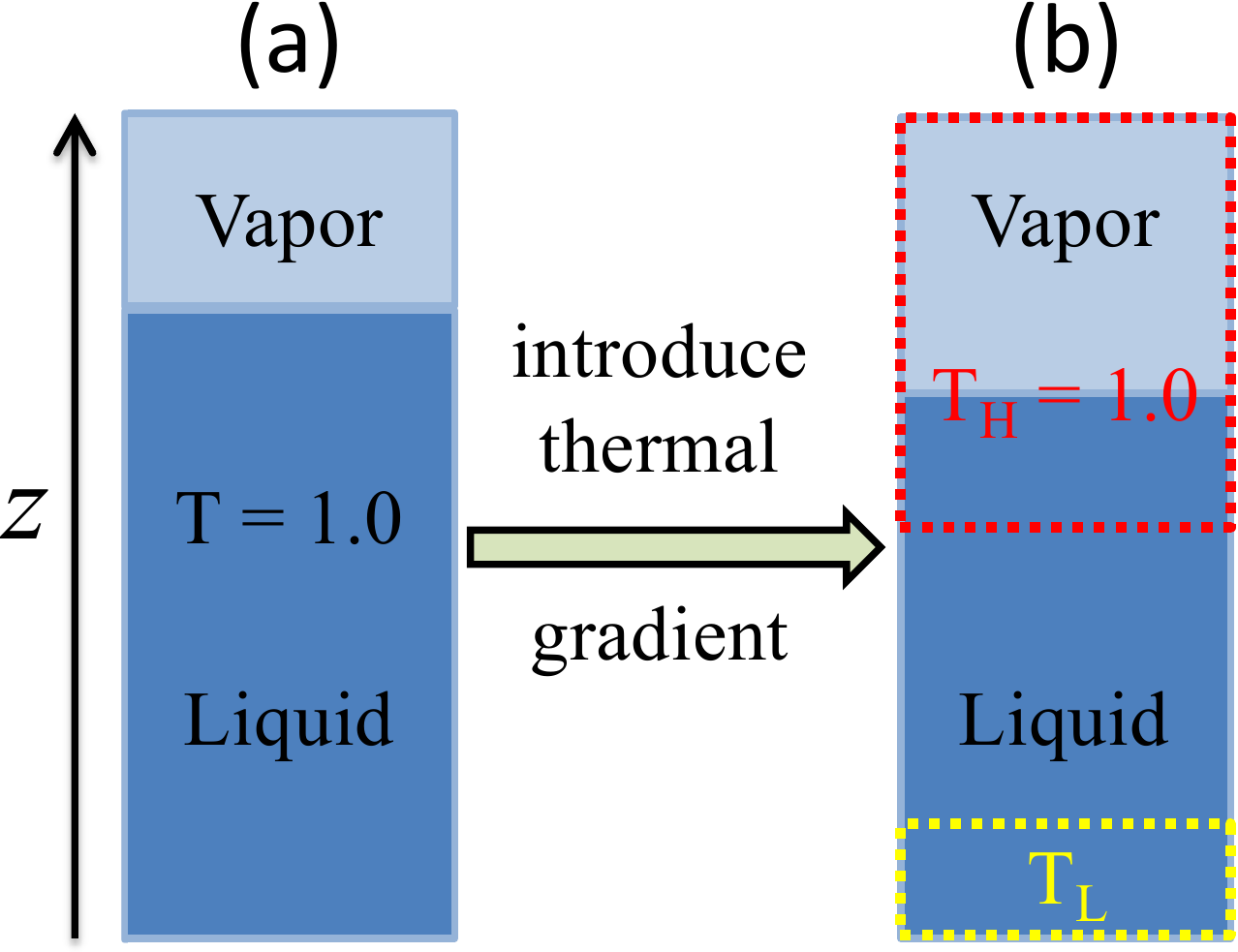}
\caption{The simulation set-up for studying thermophoresis: (a) the entire solvent and vapor are thermalized at $T=1.0\epsilon/k_\text{B}$; (b) a top region of the liquid solvent and all vapor are thermalized at $T_H=1.0\epsilon/k_\text{B}$ while a layer of the solvent adjacent to the bottom wall is thermalized at $T_L$. A positive thermal gradient is introduced into the system along the $z$-axis by using $T_L = T_H-0.1\epsilon/k_\text{B}$ or $T_L = T_H-0.3\epsilon/k_\text{B}$.}
\label{fg:T_gradients}
\end{figure}

\begin{figure}[htp]
\includegraphics[width = 0.5\textwidth]{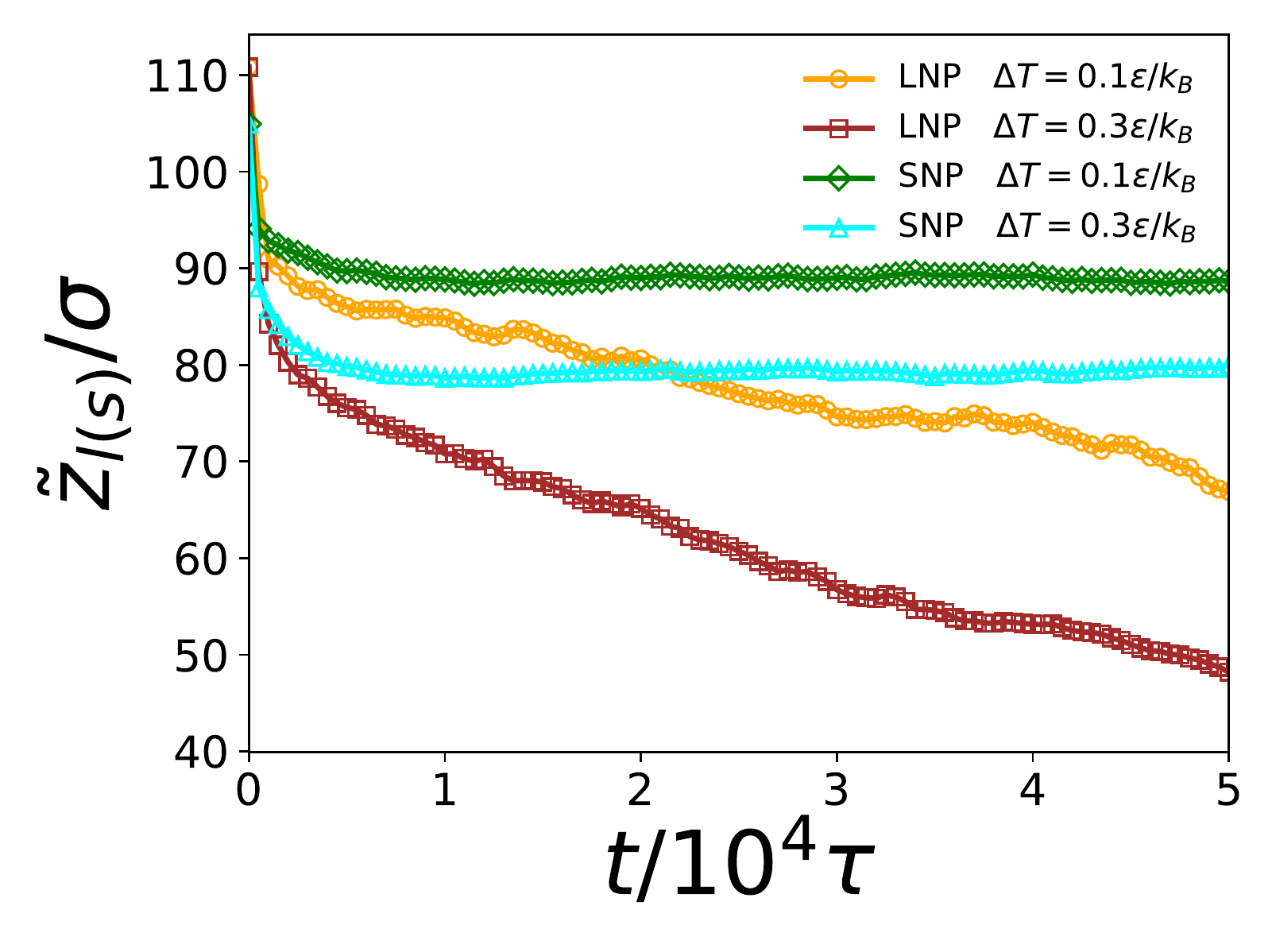}
\caption{Average position of LNPs (orange circles and dark brown squares) and SNPs (green diamonds and cyan triangles) as a function of time after a positive thermal gradient along the $z$-axis is introduced into the system as described in Fig.~\ref{fg:T_gradients}. The data are for $\Delta T \equiv T_H-T_L=0.3\epsilon/k_\text{B}$ (dark brown squares and cyan triangles) and $0.1\epsilon/k_\text{B}$ (orange circles and green diamonds).}
\label{fg:thermophoresis}
\end{figure}

\noindent To understand the thermophoresis of nanoparticles, we performed additional simulations for suspensions of only SNPs or only LNPs at volume fractions close to those in the suspension of the mixture of SNPs and LNPs. Each suspension was first equilibrated at $T=1.0\epsilon/k_\text{B}$. Then a thermal gradient was introduced into the system by thermalizing a top region of the liquid solvent and all vapor at $T_H=1.0\epsilon/k_\text{B}$ while thermalizing a layer of the solvent adjacent to the bottom wall at $T_L$, as shown in Fig.~\ref{fg:T_gradients}. Two values of $T_L$, $0.9\epsilon/k_\text{B}$ and $0.7\epsilon/k_\text{B}$, were used to generate a thermal gradient with different magnitudes in the direction perpendicular to the liquid-vapor interface. The average position of nanoparticles in each system is plotted in Fig.~\ref{fg:thermophoresis}. Since $T_L$ is lower than the initial temperature of the equilibrium suspension, the liquid contracts and the liquid-vapor interface recedes when the thermal gradient is imposed. Our data show that for the nanoparticle-solvent interaction with $A_\text{ns} = 100\epsilon$, the SNPs first move toward the substrate because of the contraction of the liquid solvent. After this transient phase, the SNPs do not show any response to the imposed thermal gradient and their average position remains almost constant with time. However, the LNPs show a clear thermophoretic response to the thermal gradient and drift toward the cooler region where the liquid density is higher. These independent simulations demonstrate that for the parameters used in this paper, the LNPs exhibit strong thermophoresis while the SNPs exhibit no thermophoresis.

Thermophoresis can be characterized by the Soret coefficient $S_T \equiv D_T/D$, where $D_T$ is the particle's thermal diffusion coefficient and $D$ is its diffusion coefficient. We can estimate $D_T$ from $v_T=-D_T\nabla T$, where $v_T$ is the drift velocity of the particles under the given thermal gradient $\nabla T$. For a temperature difference of $0.3 \epsilon/k_\text{B}$ imposed on a liquid film with thickness about $100\sigma$, $\nabla T \simeq 3\times 10^{-3}\epsilon/(k_\text{B}\sigma)$. The LNPs exhibit positive thermophoresis and migrate about $30\sigma$ by average toward the cooler region during $5\times 10^4\tau$ (the dark brown curve in Fig.~\ref{fg:thermophoresis}). The drift velocity $v_T$ is about $-6\times 10^{-4}\sigma/\tau$. The thermal diffusion coefficient of the LNPs is thus $D_T = -v_T/\nabla T \simeq 0.2\sigma^2 k_\text{B}/(\tau \epsilon)$. Since for the LNPs, the diffusivity $D = 3.61\times 10^{-3}\sigma^2/\tau$, the Soret coefficient of the LNPs is $S_T=D_T/D \simeq 55 k_\text{B}/\epsilon$. If we take $\epsilon/k_\text{B}\simeq 550$~K as for water, then $S_T \simeq 0.1~\text{K}^{-1}$ for the LNPs. For a temperature difference $0.1 \epsilon/k_\text{B}$ across the film (the orange curve in Fig.~\ref{fg:thermophoresis}), a similar analysis yields $S_T \simeq 110 k_\text{B}/\epsilon \simeq 0.2~\text{K}^{-1}$. Therefore, $S_T$ is at the order of $0.1~\text{K}^{-1}$ for the LNPs in our simulations. Since the SNPs do not exhibit much response to an imposed thermal gradient, we can effectively take their Soret coefficient to be 0.

\noindent {\bf S3. Additional Simulations and Results: Simultaneously Imposed Thermal Gradient and Evaporation}

\begin{figure}[tp]
\includegraphics[width = 0.7\textwidth]{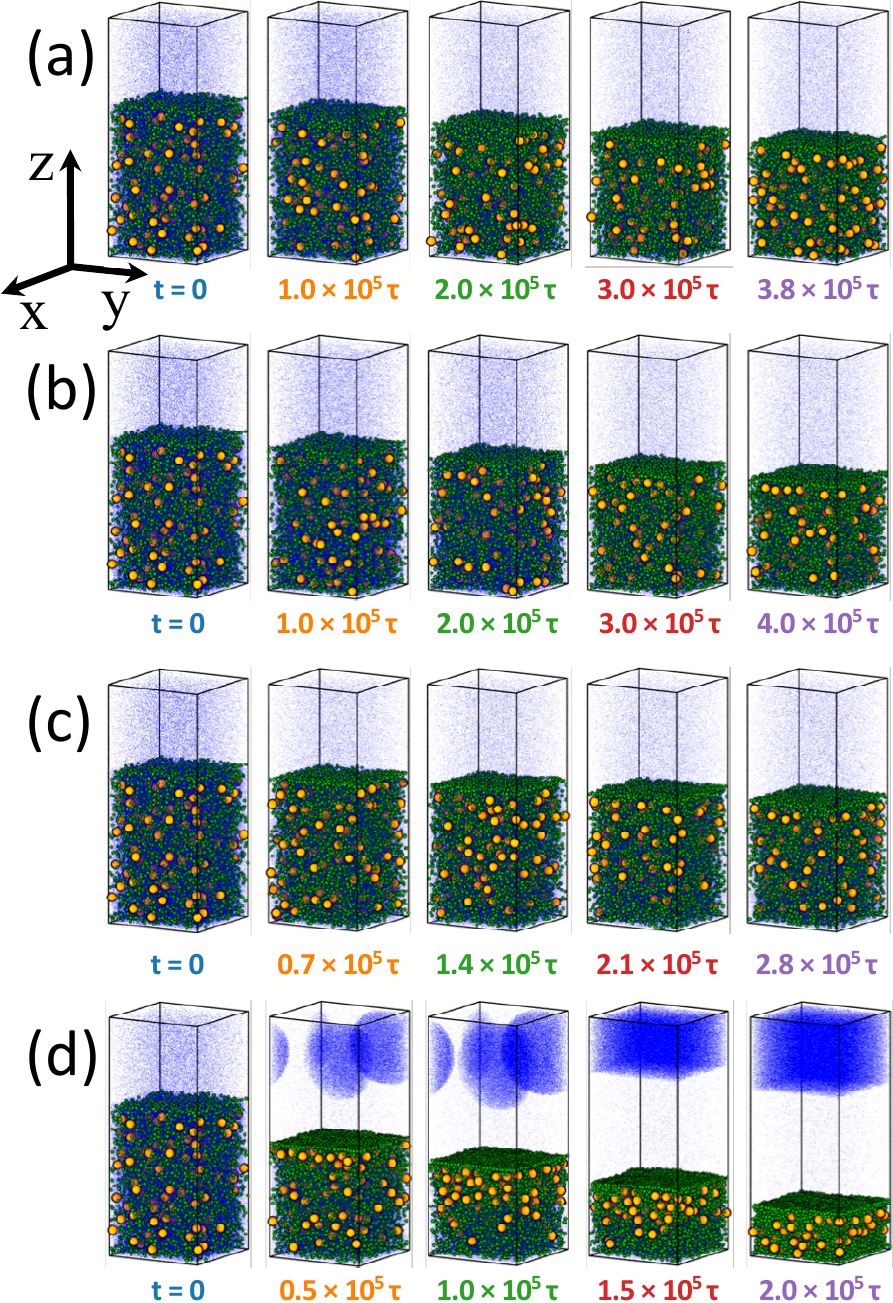}
\caption{Snapshots during solvent evaporation for (a) $T^l_{1.0}T^v_{1.1}\zeta_{5}$, (b) $T^l_{1.0}T^v_{1.05}\zeta_{5}$, (c) $T^l_{1.0}T^v_{1.0}\zeta_{5}$, and (d) $T^l_{1.0}T^v_{0.75}\zeta_{5}$. Elapsed time since evaporation was initiated at $t=0$ is listed under each snapshot. Temperature and density profiles of these systems are shown in Fig.~\ref{fg:density_SI}. Color code: SNPs (green), LNPs (orange), and solvent (blue). Only 5\% of the solvent beads are visualized to improve clarity.}
\label{fg:snaps_SI}
\end{figure}

\begin{figure}[tp]
\includegraphics[width =  \textwidth]{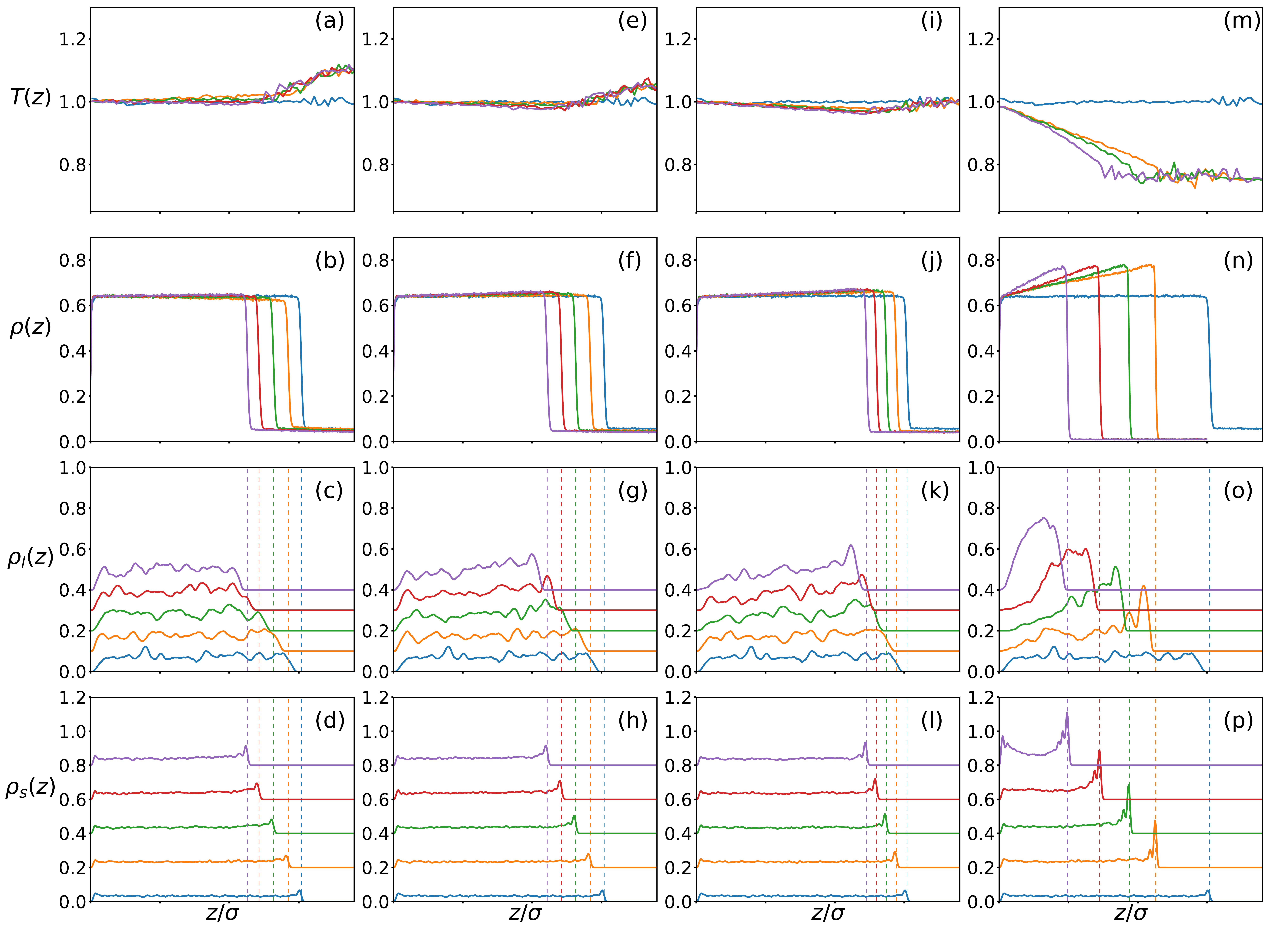}
\caption{Temperature profiles (top row) and density profiles for the solvent (second row), LNPs (third row), and SNPs (bottom row) for $T^l_{1.0}T^v_{1.1}\zeta_{5}$ (a-d), $T^l_{1.0}T^v_{1.05}\zeta_{5}$ (e-h), $T^l_{1.0}T^v_{1.0}\zeta_{5}$ (i-l), and $T^l_{1.0}T^v_{0.75}\zeta_{5}$ (m-p), respectively. The curves follow the same order as the snapshots shown in Fig.~\ref{fg:snaps_SI}. The vertical dashed lines indicate the location of the liquid-vapor interface. For clarity, the density profiles for LNPs (SNPs) are shifted upward by $0.1m/\sigma^3$ ($0.2m/\sigma^3$) successively.}
\label{fg:density_SI}
\end{figure}

\noindent We ran 6 additional simulations in which the systems were thermalized with Scheme C with $T_l = 1.0\epsilon/k_\text{B}$ and $T_v = 1.1\epsilon/k_\text{B}$, $1.05\epsilon/k_\text{B}$, $1.0\epsilon/k_\text{B}$, $0.9\epsilon/k_\text{B}$, $0.85\epsilon/k_\text{B}$, and $0.75\epsilon/k_\text{B}$ at an evaporation rate set by $\zeta = 5$, as listed in Table~1 of the main text. Snapshots at various times during evaporation are shown in Fig.~\ref{fg:snaps_SI} for $T^l_{1.0}T^v_{1.1}\zeta_{5}$, $T^l_{1.0}T^v_{1.05}\zeta_{5}$, $T^l_{1.0}T^v_{1.0}\zeta_{5}$, and $T^l_{1.0}T^v_{0.75}\zeta_{5}$. The corresponding temperature and density profiles for these 4 systems are shown in Fig.~\ref{fg:density_SI}. The results for $T^l_{1.0}T^v_{0.85}\zeta_{5}$ and $T^l_{1.0}T^v_{0.9}\zeta_{5}$ are similar to those for $T^l_{1.0}T^v_{0.75}\zeta_{5}$. 

\begin{figure}[h]
\includegraphics[width = 0.5\textwidth]{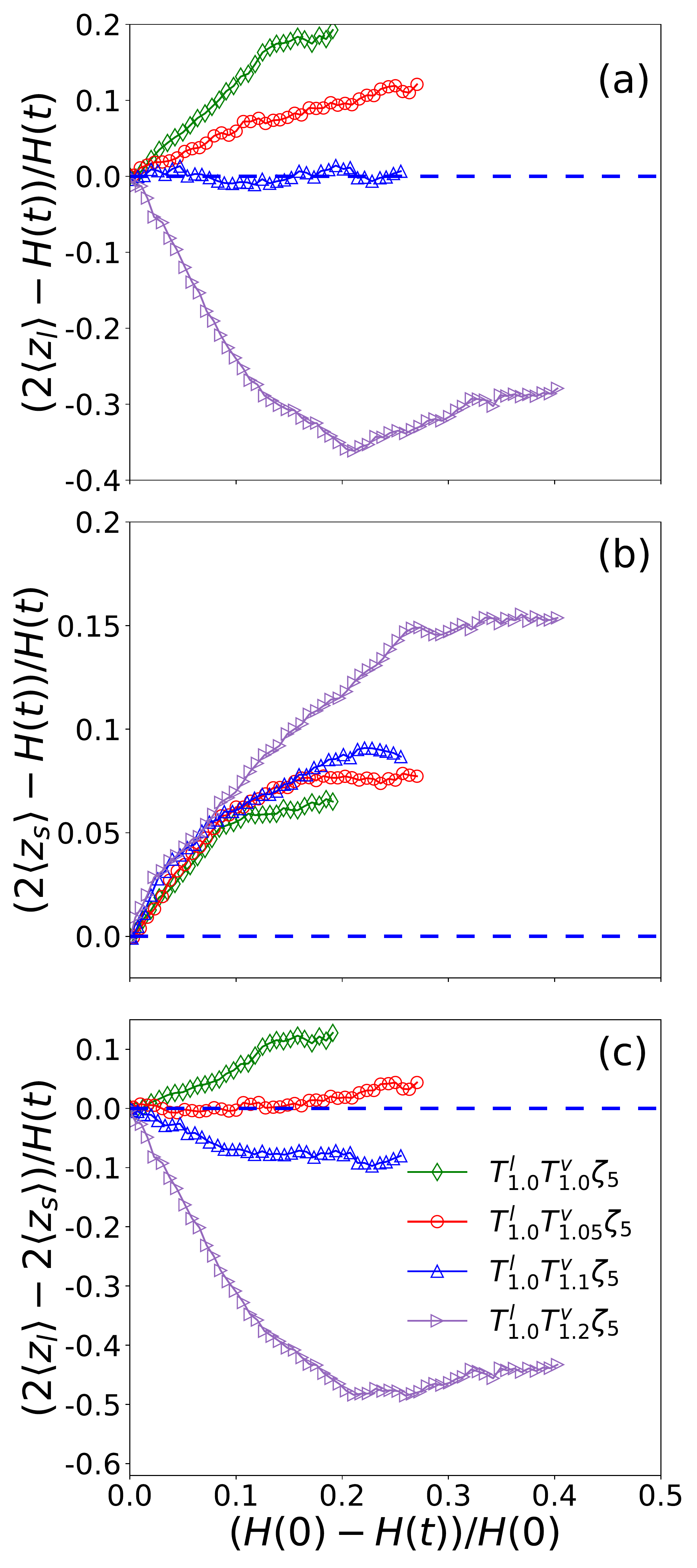}
\caption{Average position in the $z$ direction relative to the center of the film, normalized by $H(t)/2$, is plotted against the extent of drying, quantified as $(H(0)-H(t))/H(0)$, for (a) LNPs and (b) SNPs. Panel (c) shows the average separation between LNPs and SNPs, normalized by $H(t)/2$, as a function of the extent of drying. Data are for $T^l_{1.0}T^v_{1.0}\zeta_{5}$ (green diamonds), $T^l_{1.0}T^v_{1.05}\zeta_{5}$ (red circles), $T^l_{1.0}T^v_{1.1}\zeta_{5}$ (blue triangles), and $T^l_{1.0}T^v_{1.2}\zeta_{5}$ (purple right-pointing triangles).}
\label{fg:avgz_R2_SI}
\end{figure}

Similar to $T^l_{1.0}T^v_{1.2}\zeta_{5}$, $T^l_{1.0}T^v_{1.1}\zeta_{5}$ exhibits ``small-on-top'' stratification, though it is difficult to see this behavior clearly from the snapshots in Fig.~\ref{fg:snaps_SI}(a). For this system, there is a very weak positive thermal gradient in the evaporating suspension toward the liquid-vapor interface [Fig.~\ref{fg:density_SI}(a)]. Via the thermophoresis induced by this gradient, the LNPs are driven toward the substrate (i.e., toward the bottom of the liquid film) and depleted near the liquid-vapor interface. This depletion almost balances the accumulation of LNPs near the top of the film caused by the fast receding interface. As a result, the LNPs are almost uniformly distributed in the drying film, as shown in Figs.~\ref{fg:density_SI}(c) and \ref{fg:avgz_R2_SI}(a). The SNPs are essentially unaffected by the thermal gradient and accumulate below the interface since $\rm{Pe}_s > 1$ [see Figs.~\ref{fg:density_SI}(d) and \ref{fg:avgz_R2_SI}(b)]. The net outcome is ``small-on-top'' stratification for $T^l_{1.0}T^v_{1.1}\zeta_{5}$, as shown by the order parameter of stratification plotted in Fig.~\ref{fg:avgz_R2_SI}(c).

For $T^l_{1.0}T^v_{1.05}\zeta_{5}$, the evaporative cooling in the solvent is slightly stronger than the heating from the vapor thermalized at $T_v=1.05\epsilon/k_\text{B}$. The temperature profile has a very weak negative gradient in the evaporating suspension, as shown in Fig.~\ref{fg:density_SI}(e). Accordingly, the density profile of the solvent has a very weak positive gradient, as shown in Fig.~\ref{fg:density_SI}(f). Both LNPs and SNPs accumulate below the receding interface as $\rm{Pe}_l>\rm{Pe}_s > 1$ [see Figs.~\ref{fg:density_SI}(g) and \ref{fg:density_SI}(h) and Figs.~\ref{fg:avgz_R2_SI}(a) and \ref{fg:avgz_R2_SI}(b)]. The distribution of nanoparticles in this system does not stratify in the early stage of drying, though there is a transition to weak ``large-on-top'' at late times, as shown by the stratification order parameter in Fig.~\ref{fg:avgz_R2_SI}(c).

Although for $T^l_{1.0}T^v_{1.0}\zeta_{5}$ the vapor zone is coupled to a thermostat with a target temperature $1.0\epsilon/k_\text{B}$, which is equal to the temperature in the thermalized liquid layer at the bottom of the suspension, the evaporative cooling leads to a negative temperature gradient in the suspension, as shown in Fig.~\ref{fg:density_SI}(i). Consequently, the solvent density increases from its bulk value in the bottom thermalized liquid layer to a higher value near the liquid-vapor interface, as shown in Fig.~\ref{fg:density_SI}(j). The associated thermophoresis drives the LNPs toward the interfacial region [see Figs.~\ref{fg:density_SI}(k) and \ref{fg:avgz_R2_SI}(a)], though the SNPs are still accumulated right below the liquid-vapor interface as $\rm{Pe}_s > 1$ [see Figs.~\ref{fg:density_SI}(l) and \ref{fg:avgz_R2_SI}(b)]. The net outcome for $T^l_{1.0}T^v_{1.0}\zeta_{5}$ is ``large-on-top'', as shown in Fig.~\ref{fg:avgz_R2_SI}(c).

$T^l_{1.0}T^v_{1.2}\zeta_{5}$ has been discussed in detail in the main text. Fig.~\ref{fg:avgz_R2_SI} shows the results of the average position of nanoparticles normal to the interface and the order parameter of stratification for $T^l_{1.0}T^v_{1.2}\zeta_{5}$, $T^l_{1.0}T^v_{1.1}\zeta_{5}$, $T^l_{1.0}T^v_{1.05}\zeta_{5}$, and $T^l_{1.0}T^v_{1.0}\zeta_{5}$. This comparison clearly shows the transition from strong to weak ``small-on-top'' stratification, to almost no stratification, and finally to ``large-on-top'' when $T_v- T_l$ is reduced from $0.2\epsilon/k_\text{B}$ to 0.

$T^l_{1.0}\zeta_{5}$ has a weak negative thermal gradient in the evaporating suspension because of evaporative cooling. Via the associated thermophoretic process, the LNPs are pushed toward the liquid-vapor interface where the solvent density is higher. Thermophoresis overpowers diffusiophoresis where the LNPs are pushed away from the interface by the concentration gradient of the SNPs. As a result, $T^l_{1.0}\zeta_{5}$ exhibits ``large-on-top'' stratification. Motivated by this observation, we ran $T^l_{1.0}T^v_{0.75}\zeta_{5}$ that has a large $|T_v- T_l|$. Our original goal was to induce a large negative thermal gradient in the evaporating suspension at a relatively small $v_s$ such that the accompanying thermophoresis could generate strong ``large-on-top'' stratification in the drying suspension. However, as the initial vapor phase is at equilibrium with a liquid at $T_l=1.0\epsilon/k_\text{B}$, cooling the vapor down to $T_v=0.75\epsilon/k_\text{B}$ leads to droplet condensation in the vapor phase, as shown in Fig.~\ref{fg:snaps_SI}(d). Eventually, a liquid film of the solvent beads is formed at the top of the simulation cell. Because of the condensation, the vapor density near the liquid-vapor interface decreases very quickly and as a result, the solvent evaporates at a rate much higher than the target rate set by $\zeta = 5$. The actual evaporation rate in $T^l_{1.0}T^v_{0.75}\zeta_{5}$ is close to that in $T^l_{1.0}\zeta_{30}$. The temperature profile in the evaporating suspension [Fig.~\ref{fg:density_SI}(m)] and the density profiles of the solvent [Fig.~\ref{fg:density_SI}(n)] and nanoparticles [Figs.~\ref{fg:density_SI}(o) and \ref{fg:density_SI}(p)] for $T^l_{1.0}T^v_{0.75}\zeta_{5}$ are also similar to those for $T^l_{1.0}\zeta_{30}$ [see Figs.~3(a) to 3(d) of the main text]. As $T^l_{1.0}\zeta_{30}$, $T^l_{1.0}T^v_{0.75}\zeta_{5}$ exhibits ``large-on-top'' stratification.

Increasing $T_v$ to $0.85\epsilon/k_\text{B}$ or $0.90\epsilon/k_\text{B}$ yields similar droplet condensation in the vapor phase, though the extent of condensation decreases as $T_v$ is increased toward $T_l$. As a result, the receding speed of the liquid-vapor interface during evaporation decreases toward the target rate set by $\zeta = 5$.

\begin{figure}[h]
\includegraphics[width = 0.5\textwidth]{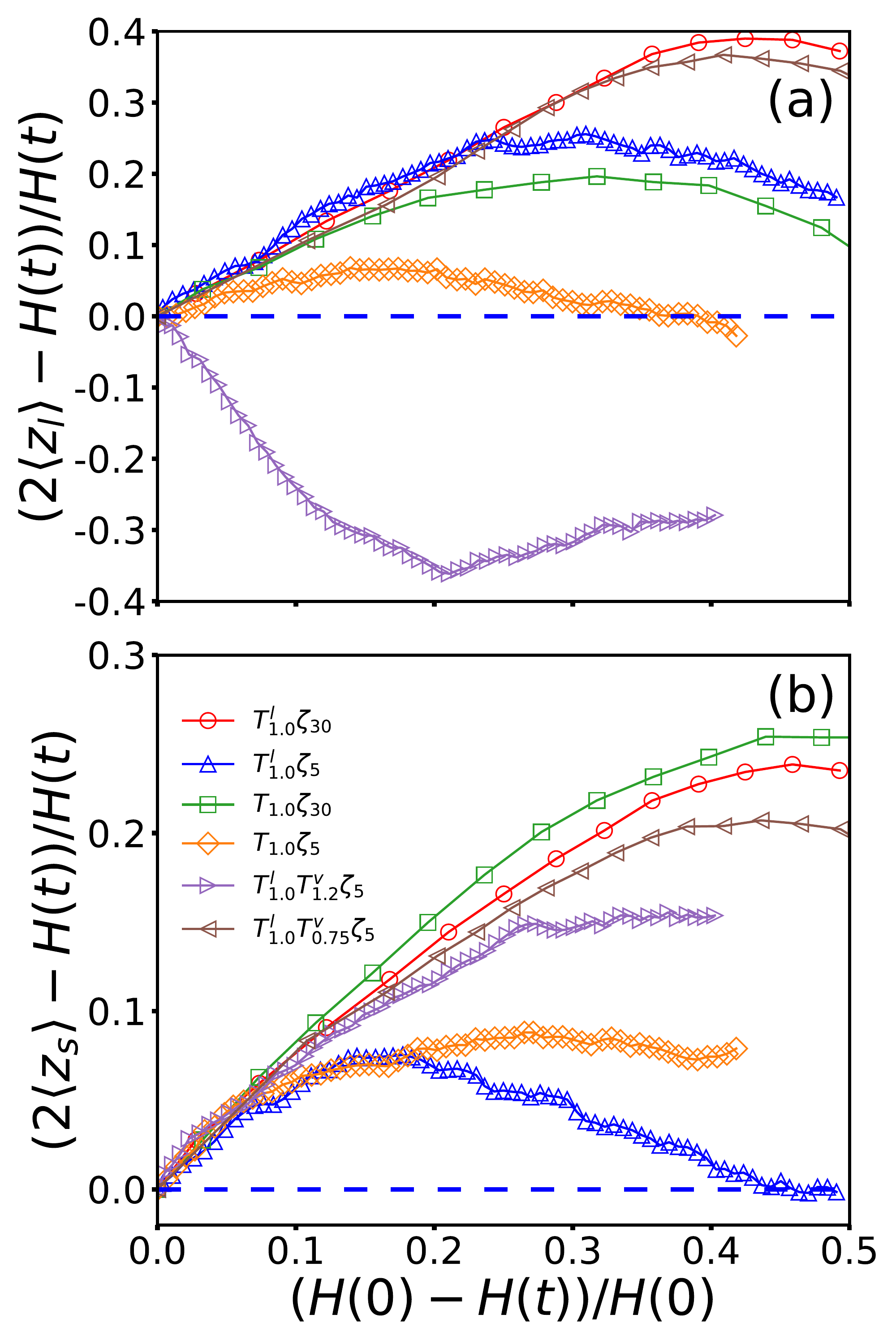}
\caption{Average position in the $z$ direction relative to the center of the film, normalized by $H(t)/2$, is plotted against the extent of drying, quantified as $(H(0)-H(t))/H(0)$, for (a) LNPs and (b) SNPs. Data are for $T^l_{1.0}\zeta_{30}$ (red circles), $T^l_{1.0}\zeta_{5}$ (blue triangles), $T_{1.0}\zeta_{30}$ (green squares), $T_{1.0}\zeta_{5}$ (orange diamonds), $T^l_{1.0}T^v_{1.2}\zeta_{5}$ (purple right-pointing triangles), and $T^l_{1.0}T^v_{0.75}\zeta_{5}$ (brown left-pointing triangles).}
\label{fg:avgz_6_sys_SI}
\end{figure}

The average position of nanoparticles during drying is plotted in Fig.~\ref{fg:avgz_6_sys_SI} as a function of the extent of drying for the 5 systems discussed in the main text (see Table 2), as well as $T^l_{1.0}T^v_{0.75}\zeta_{5}$. It is clear that the LNPs have a large thermophoretic response while the SNPs only show changes in their average position in the late stage of drying, when the thermal gradient is varied. The behavior of SNPs is predominately affected by the receding speed of the liquid-vapor interface. The variation of their average position in the late stage of drying under different thermal gradients is due to the change in the distribution of LNPs in the drying film. For example, when the LNPs are concentrated in a region, the SNPs are driven out of the same region because of crowding.

\begin{figure}[h]
\includegraphics[width = 0.5\textwidth]{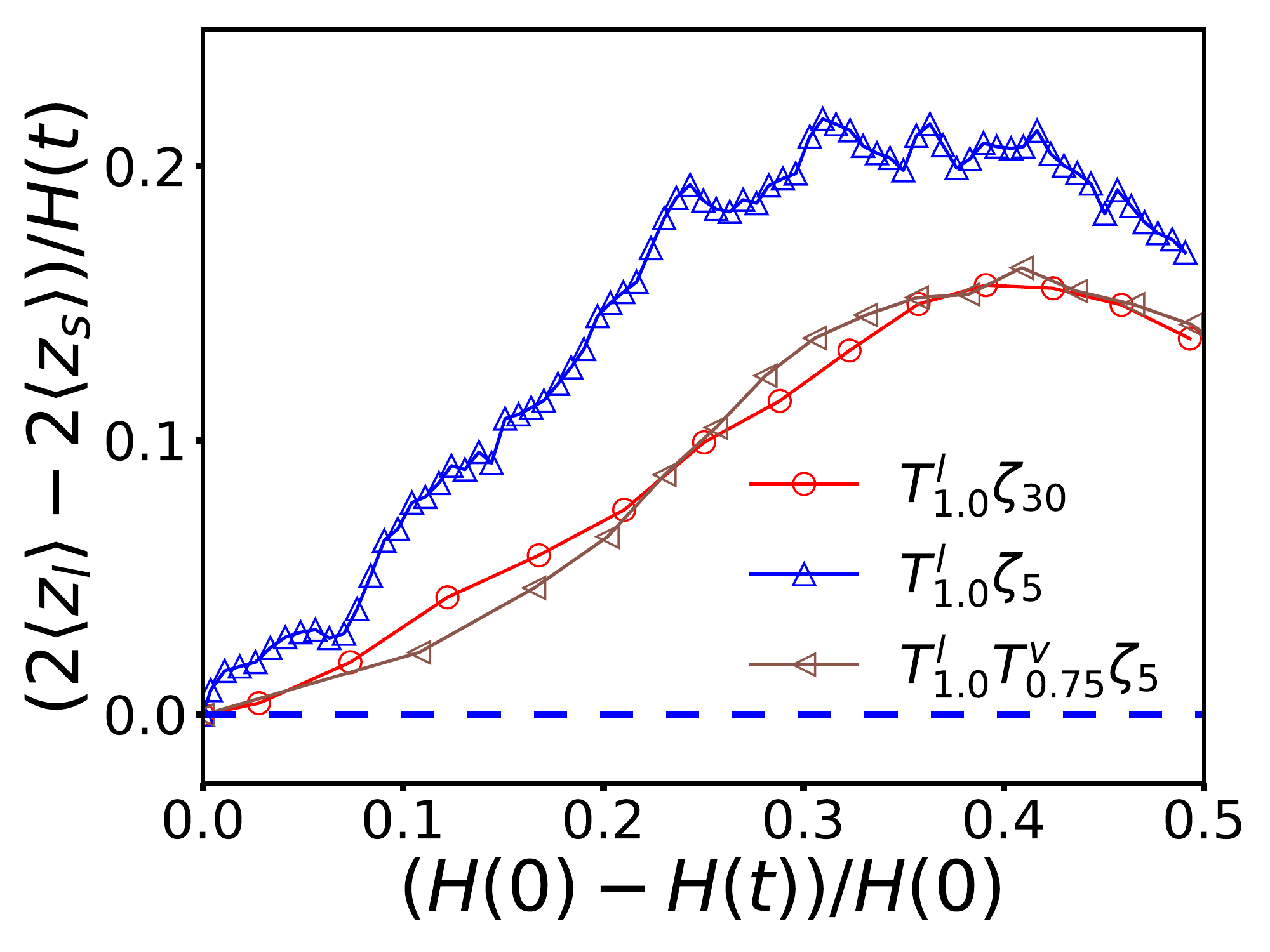}
\caption{Average separation between LNPs and SNPs, normalized by $H(t)/2$, is plotted against the extent of drying quantified as $(H(0)-H(t))/H(0)$. Data are for $T^l_{1.0}\zeta_{30}$ (red circles), $T^l_{1.0}\zeta_{5}$ (blue triangles), and $T^l_{1.0}T^v_{0.75}\zeta_{5}$ (brown left-pointing triangles).}
\label{fg:op_R3_R1_R6_SI}
\end{figure}

In Fig.~\ref{fg:op_R3_R1_R6_SI}, the order parameter of stratification is plotted as a function of the extent of drying for $T^l_{1.0}\zeta_{30}$, $T^l_{1.0}\zeta_{5}$, and $T^l_{1.0}T^v_{0.75}\zeta_{5}$. $T^l_{1.0}T^v_{0.75}\zeta_{5}$ exhibits ``large-on-top'' stratification with an amplitude very close to that in $T^l_{1.0}\zeta_{30}$ for reasons discussed previously. The reason that $T^l_{1.0}\zeta_{5}$ shows even stronger ``large-on-top'' stratification than $T^l_{1.0}\zeta_{30}$ is discussed in the main text.

\begin{figure}[h]
\includegraphics[width = 0.5\textwidth]{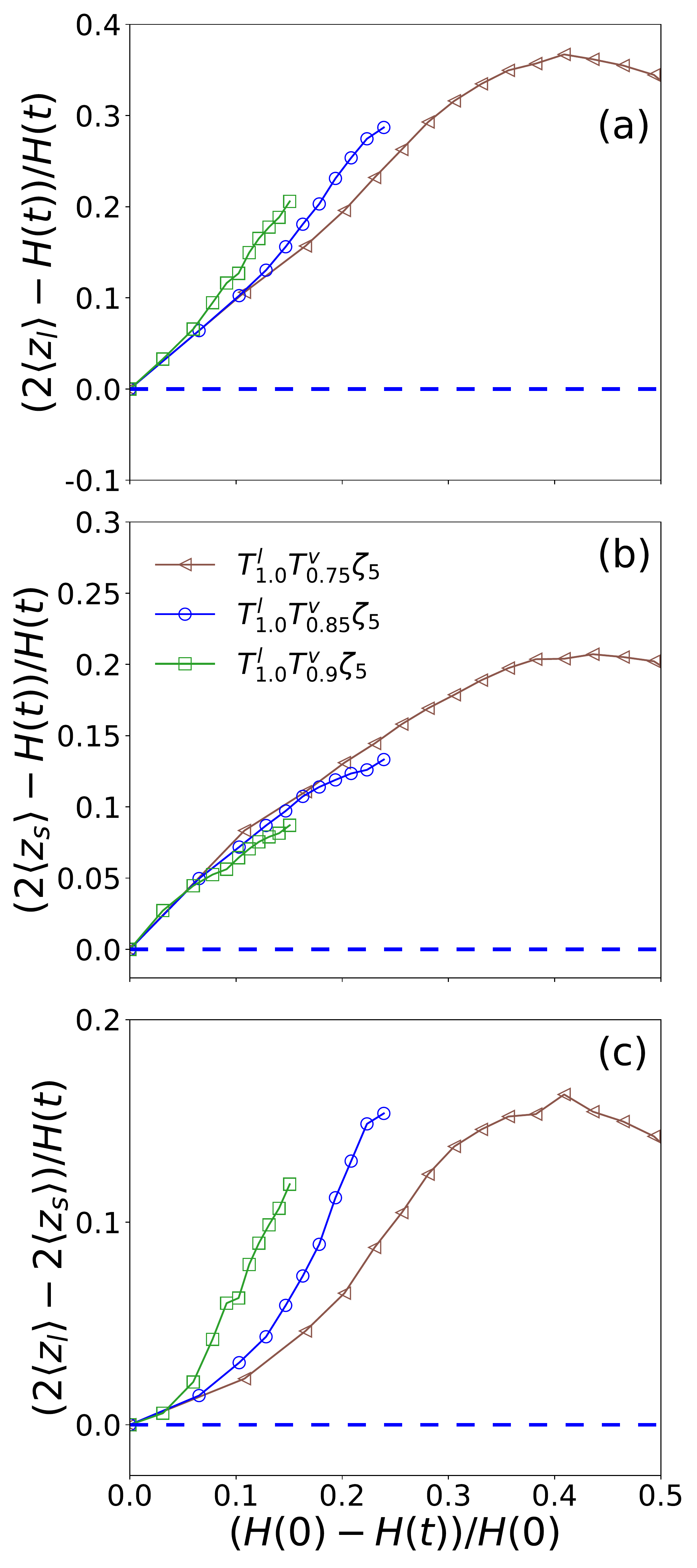}
\caption{Average position in the $z$ direction relative to the center of the film, normalized by $H(t)/2$, is plotted against the extent of drying, quantified as $(H(0)-H(t))/H(0)$, for (a) LNPs and (b) SNPs. Panel (c) shows the average separation between LNPs and SNPs, normalized by $H(t)/2$, as a function of the extent of drying. Data are for $T^l_{1.0}T^v_{0.75}\zeta_{5}$ (brown left-pointing triangles), $T^l_{1.0}T^v_{0.85}\zeta_{5}$ (blue circles), and $T^l_{1.0}T^v_{0.9}\zeta_{5}$ (green squares).}
\label{fg:avgz_R3_SI}
\end{figure}

In Fig.~\ref{fg:avgz_R3_SI}, we plot the average position of nanoparticles and the average separation between LNPs and SNPs for the 3 systems with $T_v < T_l$, i.e., $T^l_{1.0}T^v_{0.75}\zeta_{5}$, $T^l_{1.0}T^v_{0.85}\zeta_{5}$, and $T^l_{1.0}T^v_{0.9}\zeta_{5}$. While the SNPs show very little response to a thermal gradient, the distribution of LNPs is sensitive to the strength of the thermal gradient. When $T_v$ approaches $T_l$ from below, the negative thermal gradient in the evaporating suspension becomes smaller in magnitude and the driving force for LNPs to migrate to the interfacial region decreases. However, the receding speed of the liquid-vapor interface also decreases as $T_v$ is increased toward $T_l$, and the diffusiophoretic driving force that pushes LNPs way from the interfacial region thus decreases too. The complicated interplay of these two factors makes ``large-on-top'' stratification stronger when $T_v$ is increased toward $T_l$ as in $T^l_{1.0}T^v_{0.75}\zeta_{5}\rightarrow T^l_{1.0}T^v_{0.85}\zeta_{5} \rightarrow T^l_{1.0}T^v_{0.9}\zeta_{5}$ [see Fig.~\ref{fg:avgz_R3_SI}(c)].

\noindent {\bf S4. Additional Simulations and Results: Evaporation Imposed after Relaxation under a Given Thermal Gradient}

In all simulations that have been discussed so far in the main text and this Supporting Information, the evaporation process and the thermal gradient are imposed simultaneously. All these systems start from a state in which the nanoparticles are uniformly dispersed in the suspension prior to evaporation and thermophoresis. In this way, the interplay between evaporation-induced nanoparticle migration and thermophoretic motion is investigated. In this section, we report an additional simulation in which the thermal gradient is first imposed on the nanoparticle suspension. The system is relaxed under the given thermal gradient. Then the solvent is evaporated out of the relaxed system.

\begin{figure}[tp]
\includegraphics[width = 0.7\textwidth]{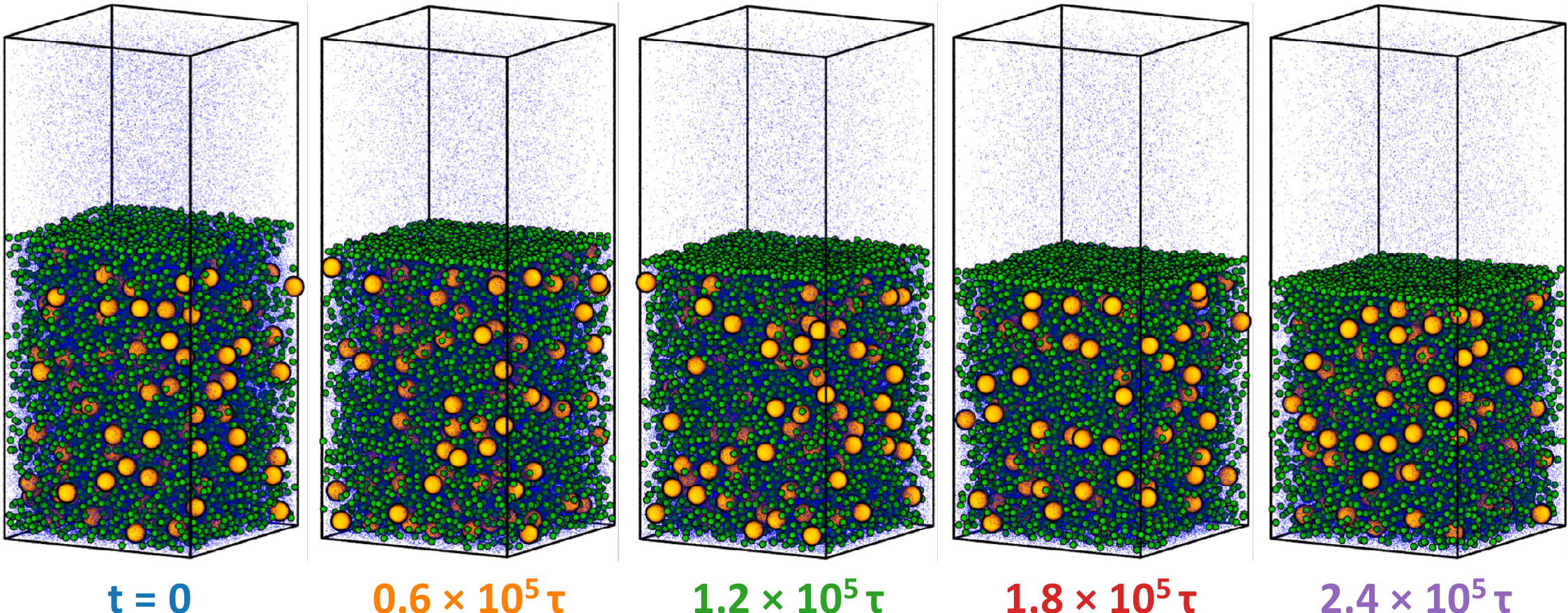}
\caption{Snapshots during solvent evaporation for $RT^l_{0.9}T^v_{1.0}\zeta_{5}$. Elapsed time since evaporation was initiated at $t=0$ is listed under each snapshot. Temperature and density profiles of this system are shown in Fig.~\ref{fg:densityR7_SI}. Color code: SNPs (green), LNPs (orange), and solvent (blue). Only 5\% of the solvent beads are visualized to improve clarity. The system was first relaxed under the imposed thermal gradient and then the relaxed state was used for the evaporation study. The time when evaporation was started is designated as $t=0$.}
\label{fg:snapsR7_SI}
\end{figure}

\begin{figure}[tp]
\includegraphics[width = 0.37 \textwidth]{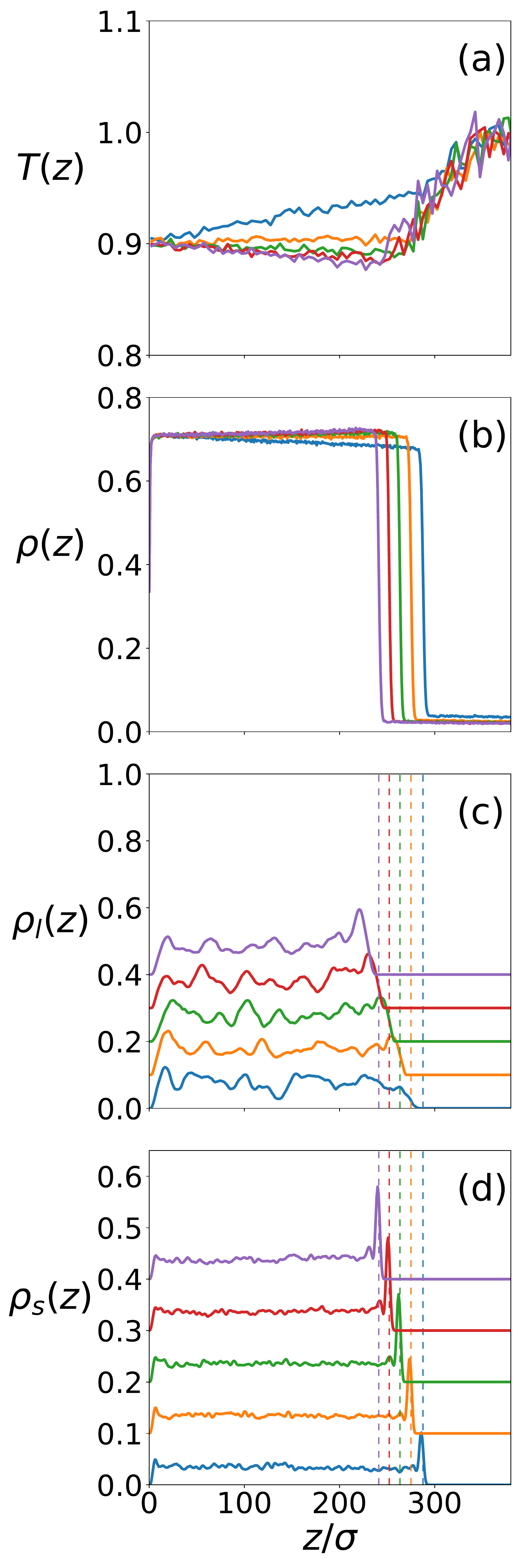}
\caption{(a) Temperature profiles and (b-d) density profiles for (b) solvent, (c) LNPs, and (d) SNPs for $RT^l_{0.9}T^v_{1.0}\zeta_{5}$. The curves follow the same order as the snapshots shown in Fig.~\ref{fg:snapsR7_SI}. The vertical dashed lines indicate the location of the liquid-vapor interface. For clarity, the density profiles for nanoparticles are shifted upward by $0.1m/\sigma^3$ successively.}
\label{fg:densityR7_SI}
\end{figure}

\begin{figure}[h]
\includegraphics[width = 0.5\textwidth]{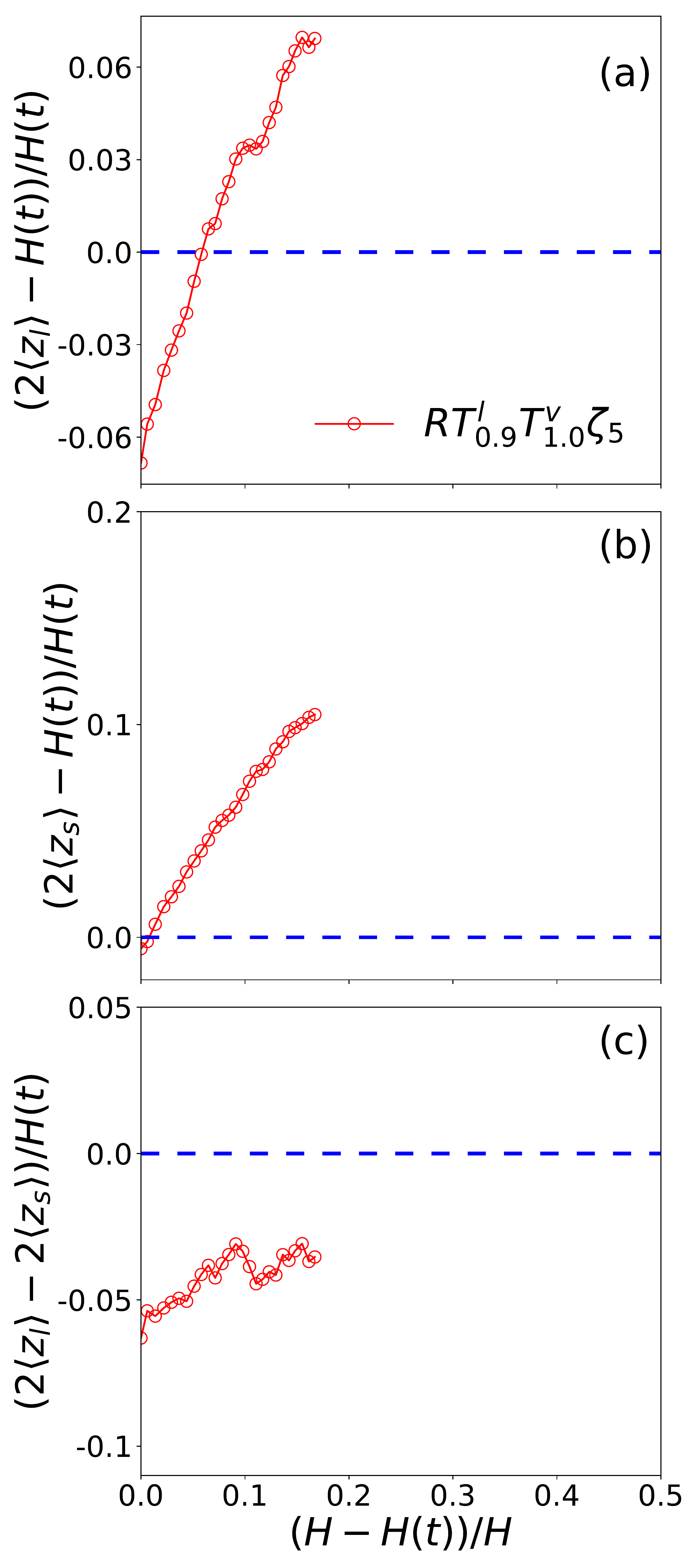}
\caption{Average position in the $z$ direction relative to the center of the film, normalized by $H(t)/2$, is plotted against the extent of drying, quantified as $(H(0)-H(t))/H(0)$, for (a) LNPs and (b) SNPs. Panel (c) shows the average separation between LNPs and SNPs, normalized by $H(t)/2$, as a function of the extent of drying. Data are for $RT^l_{0.9}T^v_{1.0}\zeta_{5}$.}
\label{fg:avgz_R7_SI}
\end{figure}

The system is designated as $RT^l_{0.9}T^v_{1.0}\zeta_{5}$, which is thermalized under Scheme C with $T_l = 0.9 \epsilon/k_\text{B}$ and $T_v = 1.0 \epsilon/k_\text{B}$, both well below the critical temperature of the Lennard-Jones solvent. The letter $R$ in the system label indicates that the system is first relaxed under the imposed thermal gradient. After relaxation, the solvent is evaporated at a rate $\zeta = 5$. In Fig.~\ref{fg:snapsR7_SI}, the snapshots of $RT^l_{0.9}T^v_{1.0}\zeta_{5}$ are shown. The snapshot at $t=0$ in Fig.~\ref{fg:snapsR7_SI} is for the relaxed system under the imposed thermal gradient. This gradient is apparent in the temperature profile (the blue curve) shown in Fig.~\ref{fg:densityR7_SI}(a). The solvent density develops a negative gradient and decreases toward the liquid-vapor interface, which can be easily seen from the density profile (the blue curve) in Fig.~\ref{fg:densityR7_SI}(b). Because of thermophoresis discussed in detail in Sec.~S2, the LNPs by average migrate toward the lower part of the suspension where temperature is lower. This thermophoretic migration is difficult to see from the snapshot at $t=0$ in Fig.~\ref{fg:snapsR7_SI} but is confirmed by the density profile of LNPs plotted in Fig.~\ref{fg:densityR7_SI}(c), where the blue curve is for the system at $t=0$, right before the evaporation process is initiated. The overall negative gradient of the blue curve in Fig.~\ref{fg:densityR7_SI}(c) indicates that, as the outcome of thermophoresis, the LNPs are more concentrated in the lower region of the suspension in the relaxed state at $t=0$. This result is also reflected in the average position of LNPs at $t=0$ plotted in Fig.~\ref{fg:avgz_R7_SI}(a). As the SNPs do not exhibit any thermophoretic responses, their distribution in the relaxed suspension is not affected by the imposed temperature gradient, as shown in Figs.~\ref{fg:snapsR7_SI}, \ref{fg:densityR7_SI}(d), and \ref{fg:avgz_R7_SI}(b). In terms of the nanoparticle distribution, the relaxed suspension is already in a ``small-on-top'' state, as indicated by the negative value of the stratification order parameter at $t=0$ shown in Fig.~\ref{fg:avgz_R7_SI}(c) .

During the evaporation process of $RT^l_{0.9}T^v_{1.0}\zeta_{5}$, a weak negative temperature gradient emerges in the drying suspension as shown in Fig.~\ref{fg:densityR7_SI}(a) because of evaporative cooling, even though $T_v > T_l$. Correspondingly, the solvent density develops a weak positive gradient and increases slightly near the liquid-vapor interface, as shown in Fig.~\ref{fg:densityR7_SI}(b). The outcome of thermophoresis during evaporation is therefore to drive the LNPs toward the liquid-vapor interface, which is apparent from the density profiles of LNPs shown in Fig.~\ref{fg:densityR7_SI}(c). At $t=0$ just prior to evaporation, the density profile of LNPs [the blue curve in Fig.~\ref{fg:densityR7_SI}(c)] exhibits a negative gradient as discussed previously. At $t=2.4\times 10^5\tau$ when 1.2 million solvent beads are evaporated, the density profile of LNPs [the purple curve in Fig.~\ref{fg:densityR7_SI}(c)] has developed a positive gradient, indicating that by average the LNPs migrate toward the interface during evaporation because of evaporative cooling and the associated thermophoresis. Although the SNPs do not respond to a thermal gradient, they accumulate just below the liquid-vapor interface as $\text{Pe}_s > 1$. This accumulation can be seen from the density profiles of SNPs plotted in Fig.~\ref{fg:densityR7_SI}(d).

For $RT^l_{0.9}T^v_{1.0}\zeta_{5}$, because the imposed thermal gradient is outweighed by evaporative cooling, the temperature profile in the drying suspension has a negative gradient from the bottom of the suspension to the liquid-vapor interface. The resulting thermophoresis makes the LNPs to drift toward the interface and their accumulation is even more significant than the enrichment of SNPs near the receding liquid-vapor interface. As a consequence, the ``small-on-top'' stratification in the relaxed system is weakened during drying, as shown by the plot of the stratification order parameter \textit{vs} time in Fig.~\ref{fg:avgz_R7_SI}(c).

\noindent {\bf S5. Additional Simulations and Results: DPD Thermostat}

In all simulations discussed previously, a Langevin thermostat that is weakly coupled to the solvent beads has been used to control temperature. In Scheme A and Scheme C, a layer of solvent adjacent to the bottom wall is thermalized in a drying suspension. Additionally, the vapor zone is also thermalized in Scheme C. In these schemes, the most part of the system including the liquid-vapor interface has no thermostats and the dynamics is controlled by the interactions between the particles (i.e., the Hamiltonian). Local momentum conservation is preserved away from the thermalized zones. The hydrodynamic interactions between nanoparticles are captured with this approach.

\begin{figure}[tp]
\includegraphics[width = 0.7\textwidth]{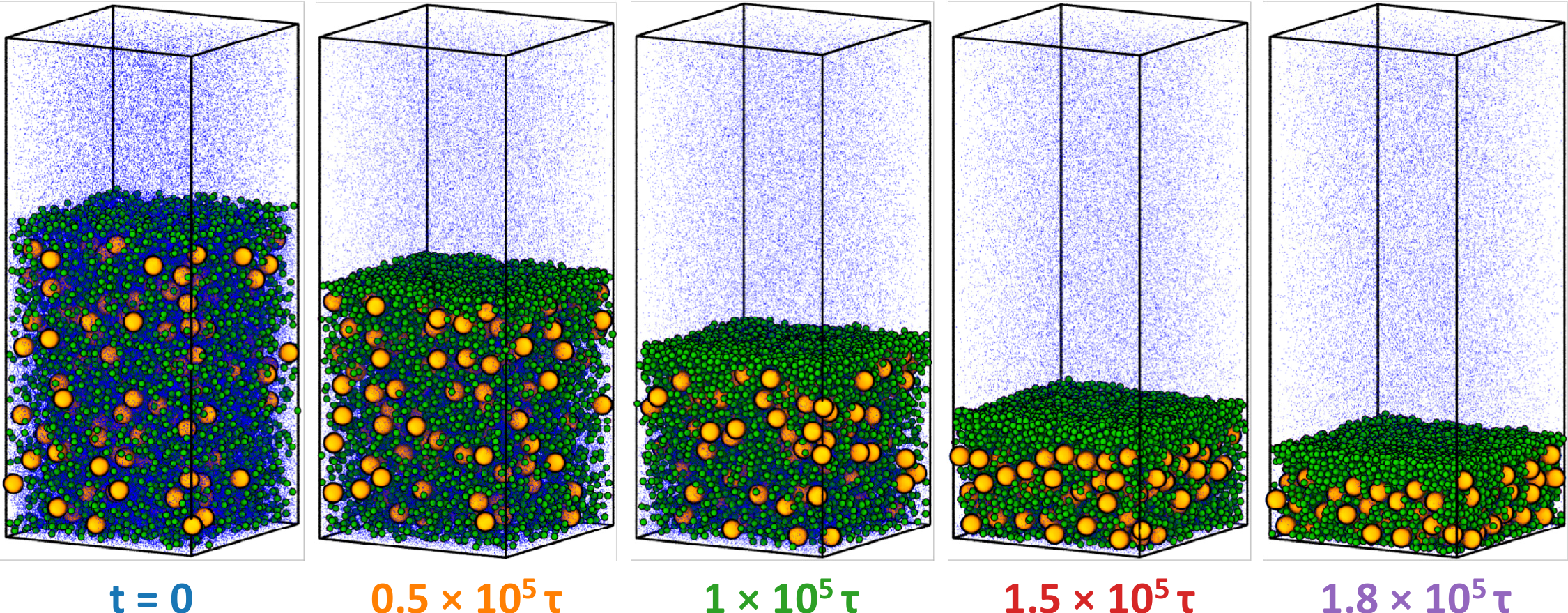}
\caption{Snapshots during solvent evaporation for $T_{1.0}\zeta_{30}$ with a DPD thermostat. Elapsed time since evaporation was initiated at $t=0$ is listed under each snapshot. Temperature and density profiles of this system are shown in Fig.~\ref{fg:densityR2_DPD_SI}. Color code: SNPs (green), LNPs (orange), and solvent (blue). Only 5\% of the solvent beads are visualized to improve clarity.}
\label{fg:snapsR2_DPD_SI}
\end{figure}

\begin{figure}[tp]
\includegraphics[width = 0.37 \textwidth]{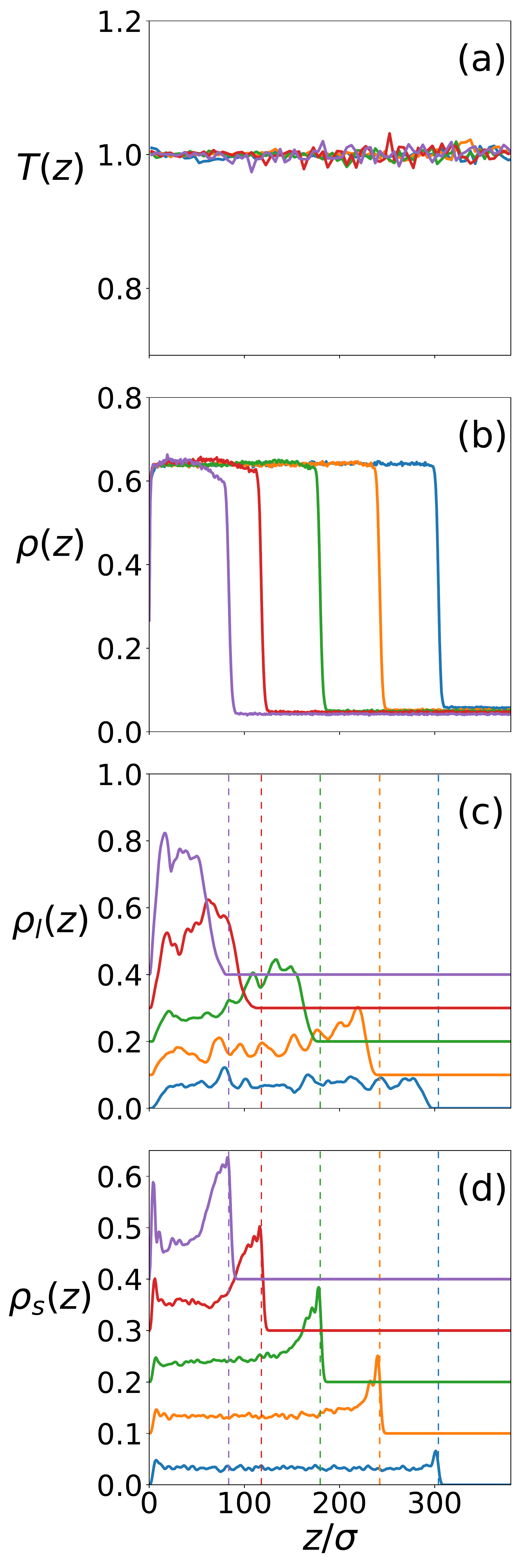}
\caption{(a) Temperature profiles and (b-d) density profiles for (b) solvent, (c) LNPs, and (d) SNPs for $T_{1.0}\zeta_{30}$ with a DPD thermostat. The curves follow the same order as the snapshots shown in Fig.~\ref{fg:snapsR2_DPD_SI}. The vertical dashed lines indicate the location of the liquid-vapor interface. For clarity, the density profiles for nanoparticles are shifted upward by $0.1m/\sigma^3$ successively.}
\label{fg:densityR2_DPD_SI}
\end{figure}

\begin{figure}[h]
\includegraphics[width = 0.5\textwidth]{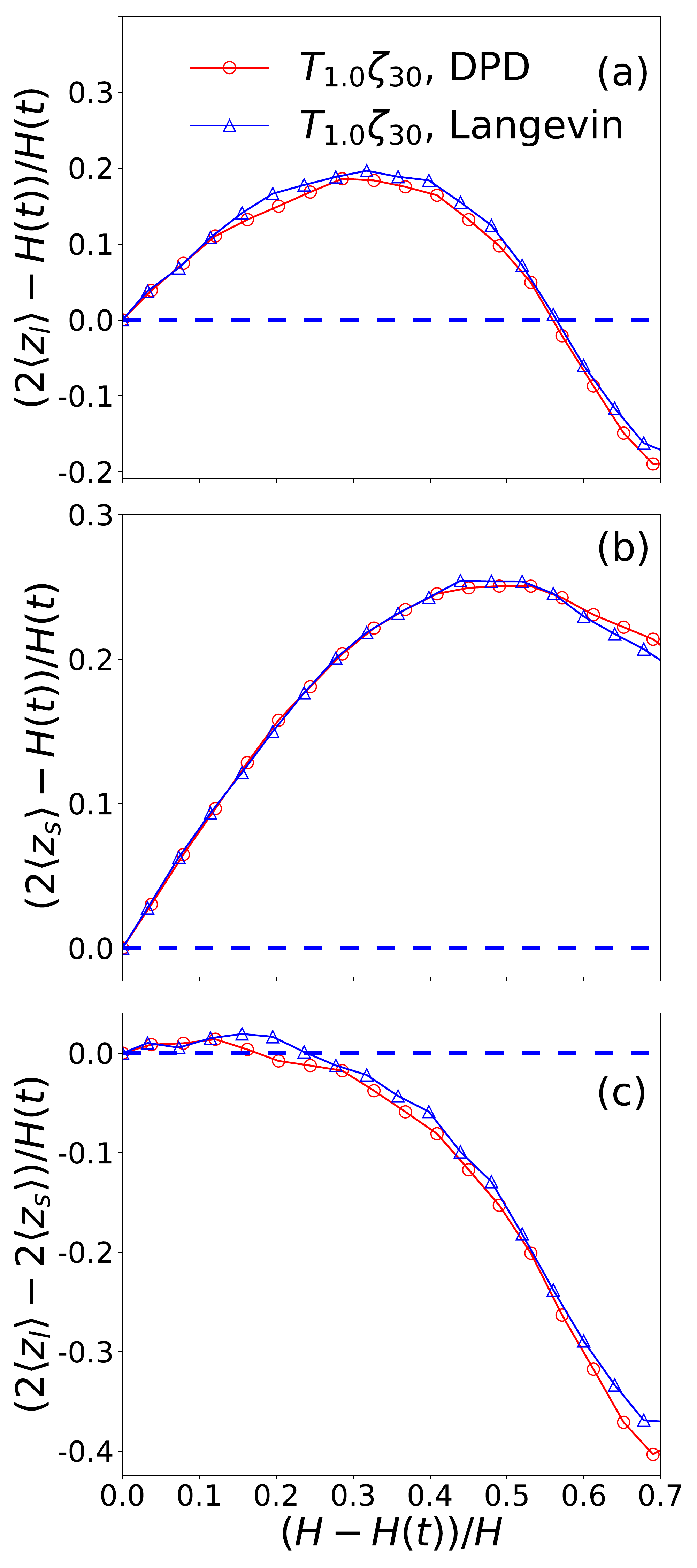}
\caption{Average position in the $z$ direction relative to the center of the film, normalized by $H(t)/2$, is plotted against the extent of drying, quantified as $(H(0)-H(t))/H(0)$, for (a) LNPs and (b) SNPs. Panel (c) shows the average separation between LNPs and SNPs, normalized by $H(t)/2$, as a function of the extent of drying. Data are for $T_{1.0}\zeta_{30}$ with a Langevin (blue triangles) and a DPD (red circles) thermostat.}
\label{fg:avgz_R2_DPD_SI}
\end{figure}

In Scheme B, all solvent beads in the system are coupled to a Langevin thermostat in order to achieve an isothermal system during evaporation. To address whether the observed phenomena including stratification may be affected by the Langevin thermostat employed in the simulation, though the strength of the thermostat is weak as we use a small damping rate, $0.01\tau^{-1}$, we performed an additional simulation in Scheme B with the temperature of the system controlled by a thermostat based on dissipative particle dynamics (DPD) [for detail, see \url{https://lammps.sandia.gov/doc/pair_dpd.html}]. In the DPD thermostat, two particles experience an equal but opposite viscous drag force that depends on the difference between the velocities of the two particles.\cite{Tsige2004b} For the entire system, all these drag forces cancel each other and the net drag is 0. The DPD thermostat is thus momentum-conserving and appropriate for capturing the hydrodynamic interactions in a suspension. We simulated the drying process of $T_{1.0}\zeta_{30}$, where the temperature is kept at $1.0\epsilon/k_\text{B}$ using a DPD thermostat with a small drag coefficient $\gamma = 0.05m/\tau$. The results, including the snapshots of the system, the temperature and density profiles, and the order parameter of stratification, are shown in Figs.~\ref{fg:snapsR2_DPD_SI}, \ref{fg:densityR2_DPD_SI}, and \ref{fg:avgz_R2_DPD_SI}, respectively. In summary, all results obtained with the DPD thermostat are very close to those obtained with the Langevin thermostat. Therefore, it is safe to use a weak Langevin thermostat to control temperature in molecular dynamics simulations of particle suspensions.

\noindent {\bf S6. Mapping between Lennard-Jones and Real Units}

All results in this paper are reported in the reduced Lennard-Jones (LJ) units based on $\epsilon$ (energy unit), $m$ (mass unit), and $\sigma$ (length unit). Here we provide details on how the rough mapping between the LJ and real units given in Table 1 of the main text is derived. The critical temperature of the LJ fluid adopted here is estimated to be about $1.18\epsilon/k_\text{B}$.\cite{Cheng2011, Watanabe2012} By mapping this temperature to the critical temperature of water, 647 K, we obtain $\epsilon/k_\text{B} \simeq 550$ K and $\epsilon \simeq 7.6\times 10^{-21}$ J. The critical pressure of the LJ fluid used here as the solvent is estimated to be around $0.12 \epsilon/\sigma^3$.\cite{Potoff1998, Caillol1998} By mapping this pressure to the critical pressure of water, 22 MPa, we obtain $\sigma \simeq 0.35\times 10^{-9}$ m. The critical density of the LJ fluid used in this paper is around $0.31 m/\sigma^3$.\cite{Cheng2011, Watanabe2012, Potoff1998, Caillol1998} By mapping this density to the critical density of water, 322 $\text{kg}/\text{m}^3$, we obtain $m\simeq 4.5\times 10^{-26}$ kg. After the mapping of $\epsilon$, $m$, and $\sigma$ to real values is determined, the mapping of the other  LJ units to the real units can be derived, which is summarized in Table 1 of the main text. However, the rough mapping only serves as a crude guidance to expedite the comparison of the results reported in this paper with experimental measurements. The mapping is not expected to be accurate as the LJ fluid model adopted here is not anticipated to be an accurate model of water. For example, in this mapping the LJ liquid density is about $0.64m/\sigma^3 \simeq 672~\text{kg}/\text{m}^3$ at temperature $1.0\epsilon/k_\text{B} \simeq 550$ K.\cite{Cheng2013} However, the water density at 550 K is about $750~\text{kg}/\text{m}^3$. The dynamic viscosity of the LJ liquid used here is about $1.0 m/(\tau\sigma) \simeq 1.5\times 10^{-4}$ Pa s at $1.0\epsilon/k_\text{B} \simeq 550$ K.\cite{Petersen2010, Cheng2012} However, the viscosity of water at 550 K is about $30\%$ lower at about $1.0\times 10^{-4}$ Pa s.



\end{document}